\documentclass[aps,prl,preprint,superscriptaddress,floatfix,noshowpacs]{revtex4-1}
\usepackage{ifthen}

\usepackage{amsmath,amssymb,amsthm}
\usepackage{calc}
\usepackage{relsize}
\usepackage{graphicx}

\makeatletter

\renewcommand\@biblabel[1]{#1.}

\newenvironment{addendum}{%
    \setlength{\parindent}{0in}%
    \small%
    \begin{list}{Acknowledgements}{%
        \setlength{\leftmargin}{0in}%
        \setlength{\listparindent}{0in}%
        \setlength{\labelsep}{0em}%
        \setlength{\labelwidth}{0in}%
        \setlength{\itemsep}{12pt}%
        }
    }
    {\end{list}\normalsize}

\makeatother

\newcommand{\emphas}[1]{\textit{#1}}
\newcommand{\emphb}[1]{#1}
\newcommand{\ignr}[1]{}

\DeclareMathOperator{\Thrm}{\textrm{MC}_{\mbox{\ensuremath{\scriptstyle{E}}}}}
\newcommand{\indx}{\nu}
\newcommand{\Npcn}{\left(N_{\!\textrm{pc}}\right)_{\alpha}^{\{\vec{n}\}}}
\newcommand{\Npca}{N_{\textrm{pc}}}
\newcommand{\etaa}{\eta^{\{\!\alpha\}}_{\vec{n}}}
\newcommand{\etaMC}{\eta^{\{\!\alpha\}}_{\;\textrm{MC}\{\!\vec{n}\}}}
\newcommand{\NinMC}{N}

\makeatletter
\newlength{\alength@}
\newlength{\widtha@}
\newlength{\widthb@}
\newlength{\widthc@}
\newlength{\widthd@}
\newlength{\heighta@}
\newlength{\heightb@}
\newlength{\heightc@}
\newlength{\heightd@}
\newlength{\heighte@}
\newcommand{\lenthpt}[1]{
\setlength{\alength@}{#1 * \real{0.015258789}}
\number\alength@}
\newcommand{\fintave}[2]{
\setlength{\widtha@}{\widthof{#1}}
\setlength{\heighta@}{0.5pt}
\setlength{\heighte@}{\heighta@}
\setlength{\widthb@}{\widthof{\ensuremath{\,#2}}}
\setlength{\widthc@}{1\widtha@}
\setlength{\widthd@}{\widthb@+0.5\widthc@-0.5\widtha@}
\setlength{\heightb@}{\heightof{\ensuremath{#2}}}
\setlength{\heightc@}{0.4\heightb@-0.5\heighta@}
\setlength{\heightd@}{0\heighta@}
\hspace*{-\widthd@}\stackrel{\hspace*{\widthb@}
\stackrel{\vrule height\heighta@ width\widthc@ depth0pt}{}
\,#2}{#1}
}
\makeatother
\newcommand{\tmax}{\tau}
\newcommand{\avetmax}[1]{\fintave{\ensuremath{#1}}{\ensuremath{\tmax}}}
\newcommand{\inftave}[1]{\fintave{\ensuremath{#1}}{\mathsmaller{\ensuremath{\infty}}}}
\newcommand{\init}[1]{\fintave{\ensuremath{#1}}{\mathsmaller{\ensuremath{0}}}}
\newcommand{\MC}[1]{{#1}_{\text{MC}}}

\newcommand{\bfpanel}[1]{\textbf{\MakeLowercase{#1}}}
\newcommand{\mfpanel}[1]{\MakeLowercase{#1}}
\newcommand{\IntroPanlSymb}[1]{#1,}

\newcommand{\citen}[1]{\cite{#1}}

\newcommand{\SISectionIIaModelSupporting}[1]{S1}
\newcommand{\SISectionIIproof}[1]{S2}
\newcommand{\secMethodsIINumericalIImodelsIIused}[1]{first}
\newcommand{\secMethodsIIRelevantIItheoreticalIIresults}[1]{second}
\newcommand{\SIfIISUPPIIHzeroIIrandomization}[1]{S2}

\begin{document}

\title{An Exactly Solvable Model for the Integrability-Chaos Transition in Rough Quantum Billiards}

\author{Maxim Olshanii}
\affiliation{Department of Physics, University of Massachusetts Boston, Boston MA 02125, USA}
\email{Maxim.Olchanyi@umb.edu}
\author{Kurt Jacobs}
\affiliation{Department of Physics, University of Massachusetts Boston, Boston MA 02125, USA}
\author{Marcos Rigol}
\affiliation{Department of Physics, Georgetown University, Washington, DC 20057, USA}
\author{Vanja Dunjko}
\affiliation{Department of Physics, University of Massachusetts Boston, Boston MA 02125, USA}
\author{Harry Kennard}
\affiliation{Cavendish Laboratory, University of Cambridge, Cambridge CB3 0HE, UK}
\affiliation{Department of Physics, University of Massachusetts Boston, Boston MA 02125, USA}
\author{Vladimir A. Yurovsky}
\affiliation{School of Chemistry, Tel Aviv University, 69978 Tel Aviv, Israel}

\begin{abstract}
A central question of dynamics, largely open in the quantum case, is to what extent it erases a system's memory of its initial properties. Here we present a simple statistically solvable quantum model describing this memory loss across an integrability-chaos transition under a perturbation obeying no selection rules. From the perspective of quantum localization-delocalization on the lattice of quantum numbers, we are dealing with a situation where every lattice site is coupled to every other site with the same strength, on average. The model also rigorously justifies a similar set of relationships recently proposed in the context of two short-range-interacting ultracold atoms in a harmonic waveguide. Application of our model to an ensemble of uncorrelated impurities on a rectangular lattice gives good agreement with ab initio numerics.
\end{abstract}

\maketitle

There are two basic prototypes of non-dissipative dynamics, corresponding to whether time propagation retains a very strong or a very weak memory of the initial conditions: integrable dynamics, where time evolution conserves as many independent quantities as there are degrees of freedom; and completely chaotic dynamics, where all the traces of the initial state of the system---except for its energy and perhaps a very few other quantities---are erased exponentially fast. As an integrable system is perturbed away from integrability, its many conserved quantities stop being conserved, and a fundamental problem is to understand the manner and the mechanism by which that happens.

In the case of classical systems, this problem was highlighted by the FPU ``paradox'' and was central for the development of several subfields of classical mechanics, including the theories of solitons and of dynamical chaos \cite{berman2005}. The most celebrated result is the KAM theorem, which explains the gradual disappearance of quasiperiodic orbits as the perturbation increases.

In quantum mechanics, there do exist several relevant experimental \cite{kinoshita2006,hofferberth2007,Trotzky2011} and numerical \cite{rigol2009} studies, all in the context of studies of relaxation, as well as attempts to formulate a quantum KAM theorem \cite{reichl1987}; nevertheless, the problem remains largely open.

For isolated quantum-chaotic systems, the key property governing the memory of the initial state under time propagation is eigenstate thermalization, which is defined as follows: consider the values $\langle \alpha | \hat{A} | \alpha \rangle$, which are the quantum expectation values of an observable $\hat{A}$ with respect to the eigenstates $| \alpha \rangle$, with the corresponding eigenenergies $E_{\alpha}$, of the system hamiltonian $\hat{H}$. The system has the eigenstate thermalization property with respect to the observable $\hat{A}$ if, in any limit in which the density of states increases without bound, and in any microcanonical energy window, all the values $\langle \alpha | \hat{A} | \alpha \rangle$ in the window become equal to each other (here $\langle \alpha | \hat{A} | \alpha \rangle$ is said to be in a microcanonical energy window if the eigenenergy $E_{\alpha}$, corresponding to the state $| \alpha \rangle$, lies within some specified narrow energy interval). This fact was first discovered in the context of studying the transition from quantum to classical behavior, in the initiating  paper by Shnirelman \cite{shnirelman1974} and in the body of work in mathematics and mathematical physics that immediately followed (see ref.~\citen{barnett2006} and the references therein). Eigenstate thermalization is now believed to be the main explanation for why large quantum systems behave thermodynamically \cite{deutsch1991,srednicki1994,rigol2008}. In the present context, however, of greatest interest is the residual variation among the $\langle \alpha | \hat{A} | \alpha \rangle$ values within an energy window---a consequence of the fact that the density of states is not actually infinite. This variation has been linked to a physically relevant autocorrelation function \cite{feingold1986} and studied in many examples \cite{horoi1995,barnett2006,flambaum1997,georgeot1997}; however, the most relevant point for the present purposes is that the residual variation provides an upper limit on how much difference there can be between the following two quantities: the first is the infinite time average of the quantum expectation value of an observable $\hat{A}$ with respect to the time-evolving state $| \psi(t) \rangle$,
\begin{eqnarray}
\inftave{A}
\equiv && \lim_{\tmax\to\infty} \avetmax{A}
\label{MAIN_A_relax_1}
=\sum_{\alpha} \left| \langle \alpha | \psi_{\mbox{\scriptsize init.}} \rangle \right|^2 \langle \alpha | \hat{A} | \alpha \rangle
\end{eqnarray}
where
$\avetmax{A} \equiv \frac{1}{\tmax} \int_{0}^{\tmax} \! dt \, \langle \psi(t) | \hat{A} | \psi(t) \rangle$, $| \psi_{\mbox{\scriptsize init.}} \rangle$ is the initial state, and $| \psi(t) \rangle =\sum_{\alpha}\,c_{\alpha}(t)\,|\alpha\rangle $ with $c_{\alpha}(t)=\langle \alpha | \psi_{\mbox{\scriptsize init.}} \rangle\,e^{-i t E_{\alpha}/\hbar}$; the second quantity is the microcanonical average of $\langle \alpha | \hat{A} | \alpha \rangle$. The latter is simply the average of the $\langle \alpha | \hat{A} | \alpha \rangle$ values over all states $ | \alpha \rangle$ in a narrow energy window centered at $\bar{E}=\langle  \psi_{\mbox{\scriptsize init.}} |\hat{H} |  \psi_{\mbox{\scriptsize init.}} \rangle $, the mean energy of the initial state. Note that the occupation numbers $\left| \langle \alpha | \psi_{\mbox{\scriptsize init.}} \rangle \right|^2$ are non-negligible generally only for eigenstates $| \alpha \rangle$ from a narrow energy window centered at the mean energy $\bar{E}$ (see the Supplementary Discussion for ref.~\citen{rigol2008}). Now, if there is some nonzero residual variation among the values of $\langle \alpha | \hat{A} | \alpha \rangle$ within a microcanonical window, then one can pick an initial state whose occupation numbers $\left| \langle \alpha | \psi_{\mbox{\scriptsize init.}} \rangle \right|^2$ tend to be, say, larger for states $| \alpha \rangle$ whose values $\langle \alpha | \hat{A} | \alpha \rangle$ are larger. Then the infinite time average $\inftave{A}$ will be larger than the microcanonical average of the $\langle \alpha | \hat{A} | \alpha \rangle$ values. In contrast, as pointed out in refs.~\citen{deutsch1991,srednicki1994,rigol2008}, if all the values $\langle \alpha | \hat{A} | \alpha \rangle$ are the same, then $\inftave{A}$ and the microcanonical average of the $\langle \alpha | \hat{A} | \alpha \rangle$ values are the same.

In the other extreme, in the case of integrable systems, time evolution preserves the values of all quantum numbers \cite{kinoshita2006}, although it has been shown that, at least in the case of large systems, the infinite time average is able to erase all the information about the relative phases of the eigenstates composing the initial state \cite{rigol2007,calabrese2007,cazalilla2006}. Eigenstate thermalization fails for integrable systems.

A real micro- or mesoscopic physical system is usually neither completely integrable nor completely chaotic. On the other hand, it may often be described as an integrable system perturbed by an integrability-breaking potential. It is therefore of central importance to understand how the dynamics of the system changes as the strength of the perturbing potential is increased from zero to the point where all resemblance to the original integrable system is lost.

In the present work, we show that the gradual disappearance of the integrals of motion under perturbation can, for a conceptually important class of quantum systems, be completely characterized by a simple and universal relation.
Our approach is to develop an exactly solvable statistical model that captures the essence of the physics, and then test its predictions---in the full range of parameters from the integrable regime through the well-developed chaos--- against the exact time dynamics of a Deformed Random Gaussian Model\cite{kota2001}, a \v{S}eba-type billiard \cite{seba1990}, and a two-dimensional Anderson model\cite{anderson1958,billy2008,roati2008}. All three are examples of ``rough billiards,'' by which we mean that in their interiors they have scatterers that are completely decorrelated, and, as a result, the perturbation has large matrix elements between any two momentum states. Correspondingly, in our statistical model, we will be considering perturbing potentials that obey no selection rules. From the point of view of ``localization-delocalization in the space of quantum numbers'' our model leads to a lattice with extremely long-range hoppings. On the other end of the spectrum one finds the kicked rotor \cite{fishman1982_509} and the Bunimovich stadium \cite{borgonovi1996_4744}, with a Coulomb scatterer in a box as an intermediate case \cite{altshuler1997_487}. In the thermodynamic limit, our model always leads to delocalized eigenstates: the phenomena described in this article are thus mesoscopic finite-size effects. We show that in our model, the memory of the initial conditions is describable in terms of a simple analytic expression that is valid through the full range of perturbation strengths, from the integrable to the completely quantum-chaotic extremes, and that features a great deal of universality. This type of universality in the memory of the initial conditions in quantum systems was first noticed in the context of two short-range-interacting ultracold atoms in a harmonic waveguide \cite{Yurovsky2011}. In the present paper, we give another physically relevant example of a perturbation with no selection rules: \textit{a lattice billiard with uncorrelated impurities}. Here the absence of long-range correlation allows the perturbation to couple any momentum state with any other, with, on average, the same strength.

\section*{Results}
\paragraph{Formulation of the problem and intuitive analysis}
We consider the hamiltonian of an integrable quantum system with an integrability-breaking perturbation $\hat{V}$: $\hat{H} = H_{0}(\hat{\vec{n}}) + \hat{V}$. Here the integrable part of the hamiltonian is a function $H_{0}$ of all the members of the complete set of the integrals of motion, where the latter are labeled as $\hat{\vec{n}} = \left(\hat{n}_{1},\,\hat{n}_{2},\,\ldots,\,\hat{n}_{d} \right)$; $d$ is the number of the degrees of freedom. The eigenstates and eigenvalues of $H_{0}(\hat{\vec{n}})$ are $|\vec{n}\rangle$ and $E_{\vec{n}}$, respectively; the corresponding quantities for $\hat{H}$ are  $|\alpha\rangle$ and $E_{\alpha}$, as before. We assume that both the $E_{\vec{n}}$ and the $E_{\alpha}$ spectra are free of degeneracies.
The non-degenerate spectrum is a generic property of an integrable hamiltonian unless it has mutually non-commuting integrals of motion or commensurate frequencies \cite{landau1958,yuzbashyan2002}. However, because of the absence of level repulsion, the energy levels are often near-degenerate: they are distributed as if by a Poisson process, resulting in the exponential distribution of level spacings \cite{berry1977b}.

\begin{figure}[h!]
\vspace{-1in}\mbox{}
\begin{center}
\includegraphics[scale=.7]{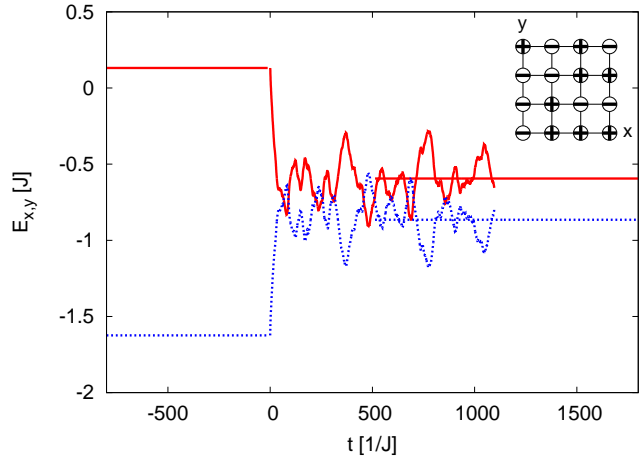}
\end{center}
\caption{
\label{f:MAIN_A_of_time}
\textbf{An example of a partially retained memory of the initial state.} Shown are the $x$- and $y$-components of the hopping energy (respectively: $E_{x}$ in solid red, with initially positive value, and $E_{y}$ in dotted blue, with initially negative value), as a function of time, of a two-dimensional $33 \times 33$-site noninteracting Anderson model in the presence of an Aharonov-Bohm (A-B) flux (see the \secMethodsIINumericalIImodelsIIused{} subsection of the Methods section). A fully chaotic systems would obey the equipartition theorem, predicting that the infinite time averages of the two components should be equal, $\inftave{(E_{x}-E_{y})}=0$. However, for intermediate strengths of the integrability-breaking perturbation, a residual deviation from equipartition---strongly correlated with $\init{(E_{x}-E_{y})}$, the initial deviation---remains. The two straight horizontal lines on the right-hand side of the plot are the exact values of $\inftave{E_{x}}$ and $\inftave{E_{y}}$, computed from equation~(\ref{MAIN_A_relax_1}). In terms of the Anderson model parameters introduced in the text following the paragraph containing equation~(\ref{MAIN_central_result}), this system has $\varepsilon = 1.0$, corresponding to the Anderson's disorder parameter of $W/J = .87$. The initial state was represented by an eigenstate of the impurity-free lattice, $ |\psi_{\mbox{\scriptsize init.}} \rangle = |n_{x}\!=\!-9,\,n_{y}\!=\!+3\rangle$, whose energy was  $E = -1.46 \, J$, where $J$ is the hopping constant (see the \secMethodsIINumericalIImodelsIIused{} subsection of the Methods section). At zero A-B flux, the ground state is $|n_{x}\!=\!0,\,n_{y}\!=\!0\rangle$. The A-B fluxes $\phi_{x}$ and $\phi_{y}$ through a complete loop along the $x$- and the $y$-directions, respectively, were chosen to be far from approximately commensurate; we had $\phi_{x} = (\varphi/4)\,\phi_{0}$ and $\phi_{y} = (e/10)\,\phi_{0}$, where $\phi_{0}$ is the elementary quantum of magnetic flux, $\varphi = 1.618\ldots$ is the golden ratio, and $e =2.718\ldots $ is the base of the natural logarithm. The apparent correlation between the two energies is due to the conservation of the total energy.
}
\end{figure}

Now we imagine that the system is prepared in some initial nonequilibrium state $|\psi_{\mbox{\scriptsize init.}}\rangle \equiv |\psi(t\!=\!0)\rangle$, and then allowed to evolve according to the hamiltonian $\hat{H}$. Following a transient period, the expectation value of a generic observable $\hat{A}$ will, overwhelming majority of the time, fluctuate around the infinite-time average given in equation~(\ref{MAIN_A_relax_1}). Figure~\ref{f:MAIN_A_of_time} gives an example of such a process, in the regime intermediate between fully integrable and fully chaotic, showing a partial retention of the memory of the initial state.

Since we are concentrating on the disappearance of the integrals of motion, we take our observable of interest to be diagonal in the eigenbasis of $H_{0}(\hat{\vec{n}})$, the integrable part of the hamiltonian:
$\langle \vec{n} | \hat{A} | \vec{n}' \rangle = A_{\vec{n}} \delta_{\vec{n}\,\vec{n}'}$.
Let us choose the initial state to be an eigenstate,
$| \vec{n}_{\mbox{\scriptsize init.}} \rangle$, of $H_{0}(\hat{\vec{n}})$.
In the intuitive discussion that follows, it will help to imagine that we chose a  state with a ``highly improbable'' (meaning: very different from the microcanonical average) value of the observable of interest $\hat{A}$.
According to equation~(\ref{MAIN_A_relax_1}), in the course of time evolution, the initial state $| \vec{n}_{\mbox{\scriptsize init.}} \rangle$ gets transformed into a superposition, with essentially random phase relationships, of the eigenstates $| \alpha \rangle$ of the perturbed hamiltonian, $\hat{H}$, such that each $| \alpha \rangle$ in the superposition has a substantial overlap with  $| \vec{n}_{\mbox{\scriptsize init.}} \rangle$, i.e. a large corresponding occupation number.

Now, since the eigenstates $| \vec{n} \rangle$ of $H_{0}(\hat{\vec{n}})$ (let us call these the ``unperturbed eigenstates'') form a basis, we may expand the eigenstates $| \alpha \rangle$ of the full, perturbed hamiltonian $\hat{H}$ (the ``perturbed eigenstates'') as linear combinations of the unperturbed eigenstates. The stronger the perturbing potential $\hat{V}$, the more different the perturbed eigenstates from the unperturbed ones, and the greater the number of unperturbed eigenstates that must be linearly combined (with appreciable weights) to build any perturbed eigenstate. A standard measure of the latter number is the so-called \emphas{number of principal components}, to be defined momentarily. Among the perturbed eigenstates  $| \alpha \rangle$ whose overlap with  $| \vec{n}_{\mbox{\scriptsize init.}} \rangle$ is non-negligible, a typical one will be such that when it is expanded in terms of the unperturbed eigenstates, the weight of
the state $| \vec{n}_{\mbox{\scriptsize init.}} \rangle$ is of the order of
$\eta_{\alpha}^{\{\vec{n}\}} \equiv \sum_{\vec{n}} \left| \langle \vec{n} | \alpha \rangle \right|^4$, which is the so-called Inverse Participation Ratio (IPR) \cite{georgeot1997}. The number of principal components mentioned above is then given by $\Npcn\sim 1/\eta_{\alpha}^{\{\vec{n}\}}$; in this case, it is the number of principal unperturbed components.

As we said, we will assume that the matrix elements $V_{\vec{n}_{1}\vec{n}_{2}}$ of the integrability-breaking perturbation $\hat{V}$ obey no selection rules. We will also assume that the equi-energy surface $H_{0}(\vec{n})$ is ``sufficiently irrational,'' which is a concept that needs a bit of explaining: as is well known, energy eigenstates in an integrable system are completely characterized by their quantum numbers (which we can take to be integers) corresponding to all the conserved quantities. In a typical integrable system, for sufficiently high excited states, if one of the quantum numbers is changed by 1 while all the others are held fixed, the change in the energy of the eigenstate will, in general, be large compared to the energy level spacing; that is, there will be many eigenstates whose energies lie in between the energy of the original state and the energy of the state that differs from the original by 1 in a single quantum number. Eigenstates that are neighbors in energy generally have very different values of all quantum numbers. In fact, if one orders the eigenstates by increasing energy and looks at how their quantum numbers change from one eigenstate to the next in the sequence, these changes may appear quite random; if they do, we say that the equi-energy surface is ``sufficiently irrational.'' More precisely: let $\{(\hat{n}_{k})_{\indx}\}$ be the sequence of the values of the $k$th quantum number as one is going from one eigenstate of $H_{0}(\vec{n})$ to the next in the order of increasing energy, labeled by $\indx$. The equi-energy surface is said to be sufficiently irrational if, for every $k$, the sequence $\{(\hat{n}_{k})_{\indx}\}$ passes any simple statistical test for randomness. This property seems to be quite generic for integrable systems with incommensurable frequencies; see Supplementary Figure~\SIfIISUPPIIHzeroIIrandomization{} for an example.

These two assumptions (the absence of selection rules in the perturbing potential $\hat{V}$ and the sufficient irrationality of the equi-energy surface of $H_{0}(\vec{n})$) imply that the presence of the state $| \vec{n}_{\mbox{\scriptsize init.}} \rangle$ as one of the components of the state $| \alpha \rangle$ does not ``bias
the selection'' of the other states $| \vec{n} \rangle$ that enter the expansion of $| \alpha \rangle$. This means that the expansion of $| \alpha \rangle$ appears as if the states (other than $| \vec{n}_{\mbox{\scriptsize init.}} \rangle$) that enter into it were indiscriminately
chosen from a microcanonical shell around $| \vec{n}_{\mbox{\scriptsize init.}} \rangle$. This would make---were it not for the ``systematic'' presence of the initial state in the expansion of $| \alpha \rangle$---the value of $\hat{A}$, on average, equal to the microcanonical average. The state $| \vec{n}_{\mbox{\scriptsize init.}} \rangle$ becomes \textit{the only} component of the state $| \alpha \rangle$ that ``remembers'' the initial value of $\hat{A}$. Since on average $\Npcn\sim 1/\eta$ unperturbed states enter the expansion of a perturbed state---and out of these unperturbed states that enter the expansion, usually exactly one has the initial value for the observable $\hat{A}$---we obtain the estimate that the infinite time average of $\hat{A}$ is the weighted average of the initial value and the microcanonical value of $\hat{A}$, where the weight ratio is $1:\left(\Npcn-1\right)$ in favor of the microcanonical value:
$
\inftave{A}
\approx
\eta_{\alpha}^{\{\vec{n}\}} \init{A}
+
(1 - \eta_{\alpha}^{\{\vec{n}\}}) \MC{A}
$,
where $\init{A}=\langle \vec{n}_{\mbox{\scriptsize init.}} | \hat{A} | \vec{n}_{\mbox{\scriptsize init.}} \rangle$ is the quantum expectation value of $\hat{A}$ with respect to the initial state, and $\MC{A}$ is the microcanonical average, i.e. the average over all the values $\langle \vec{n} | \hat{A} | \vec{n} \rangle$ such that the unperturbed eigenenergies of $| \vec{n} \rangle$ lie in a narrow window centered at the mean energy of the system.

\paragraph{A statistically solvable model.}
Guided by this intuitive reasoning, we have constructed a statistical model that captures the essential physics and can be solved exactly. The model has two principal ingredients. First, the unperturbed hamiltonian $\hat{H}_{0}$ is replaced by an ensemble of hamiltonians, each member of which has the same eigenenergies and eigenstates as $\hat{H}_{0}$, but which eigenstate corresponds to which eigenvalue is chosen at random. It is convenient to think of these random assignments as permuting the eigenvalues, while the eigenstates remain fixed; so each permutation $\sigma$, which sends the $N$-tuple $(1,\, 2,\, \ldots,\,N)$ to $(\sigma(1),\, \sigma(2),\, \ldots,\, \sigma(N))$, says that the $j$th largest eigenvalue corresponds to the eigenstate that in the original hamiltonian corresponded to the $\sigma^{-1}(j)$th largest eigenvalue. Second, the perturbation $\hat{V}$ is replaced by an ensemble of perturbations; the distribution of the perturbations (viewed as matrix elements between various eigenstates) is assumed to be invariant under the permutations $\sigma$ of the eigenstates. We should note that we are permuting only the unperturbed eigenstates whose unperturbed eigenvalues lie within a microcanonical energy window ${\cal W}_{\mbox{\scriptsize MC}}(E,\,\Delta E)$ centered
at the mean energy $E$ of the initial state. Also, the perturbations $\hat{V}$ are truncated so that they couple only the eigenstates within that energy window. The following result then follows (see the Supplementary Methods for the details of the derivation):
\begin{eqnarray}
\Big\langle \inftave{A} \Big\rangle_{\sigma,\,\hat{V}}
=
\frac{
\left(N-\Npca\right) \init{A}
+
N\left(\Npca-1\right) \MC{A}
     }{\Npca\left(N-1\right)}
\,\,
\nonumber
\\
{}
\label{MAIN_almost_the_central_result}
\end{eqnarray}
Here $N$ gives the number of the states in the microcanonical window ${\cal W}_{\mbox{\scriptsize MC}}(E,\,\Delta E)$, and
$\Big\langle \ldots \Big\rangle_{\sigma,\,\hat{V}}$ stands for an average over the uniformly distributed permutations $\sigma$ and the permutation-invariant-distributed perturbations $\hat{V}$. It will be convenient to introduce, for any quantity $B_{\vec{n}}$ that depends on the values of the quantum numbers $\vec{n}$, the \emphb{microcanonical average} as
\[
\Thrm[B_{\vec{n}}] \equiv \frac{1}{\NinMC} \sum_{\vec{n}:\,E_{\vec{n}} \in {\cal W}_{\mbox{\scriptsize MC}}(E,\,\Delta E)} B_{\vec{n}}
\]
where $\NinMC$ is the number of unperturbed eigenstates whose eigenvalues are in the microcanonical energy window. Then the typical number of the (interacting) principal components is given by
$
\Npca \equiv 1/{ \Big\langle \etaMC(\sigma,\,\hat{V}) \Big\rangle_{\sigma,\,\hat{V}} }
$
where
$
\etaMC(\sigma,\,\hat{V})
\equiv
\Thrm\left[\etaa(\sigma,\,\hat{V})\right]
$
is the \emphb{microcanonical average of the inverse participation ratio} of the noninteracting eigenstates over the interacting ones
$\left(\mbox{in other words,}\;
\etaa \equiv \sum_{\alpha} |\langle \alpha | \vec{n} \rangle |^4\right)$,
$\init{A}=\langle \vec{n}_{\mbox{\scriptsize init.}} | \hat{A} | \vec{n}_{\mbox{\scriptsize init.}} \rangle$ is the value of $\hat{A}$ in the initial state (it is written as an expectation value even though, given our choices, the initial state is also an eigenstate of $\hat{A}$), and
$
\MC{A} \equiv \Thrm\left[A_{\vec{n}}\right]
$
is the \emphb{microcanonical average for the observable}.

\begin{figure}[h!]
\begin{center}
\includegraphics[scale=.5]{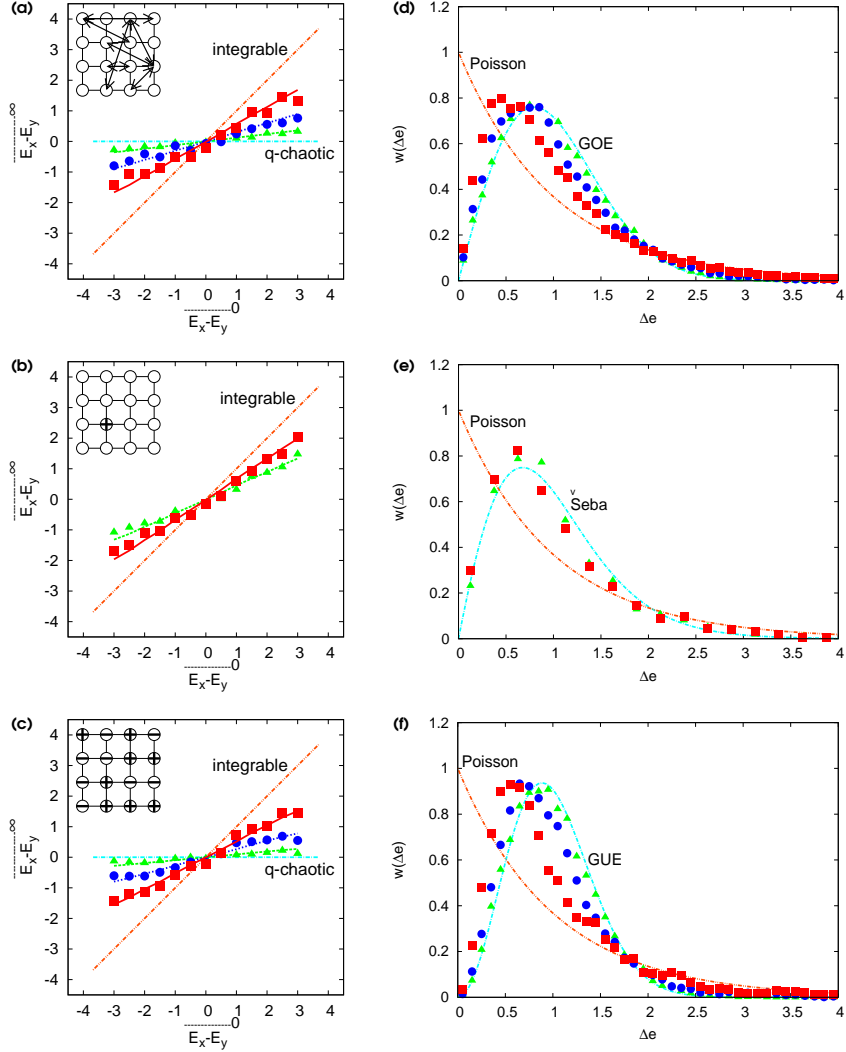}
\end{center}
\caption
{
\label{f:MAIN_ipr_and_level_statistics}
%
%
\smaller\smaller
\textbf{Testing the predictive power of equation~(\ref{MAIN_central_result}).} Our integrable system is a single particle on a two-dimensional lattice with an Aharonov-Bohm flux. Lattice parameters are the same as for Figure~\ref{f:MAIN_A_of_time}. Each row of plots corresponds to a different type of non-integrable perturbation, as follows. \IntroPanlSymb{\bfpanel{a}}\IntroPanlSymb{\bfpanel{d}} A real Gaussian random matrix perturbation acting between the eigenstates of the unperturbed lattice. \IntroPanlSymb{\bfpanel{b}}\IntroPanlSymb{\bfpanel{e}} A single impurity of fixed strength. \IntroPanlSymb{\bfpanel{c}}\IntroPanlSymb{\bfpanel{f}} An Anderson-type disorder. In plots \IntroPanlSymb{\bfpanel{a}}\IntroPanlSymb{\bfpanel{d}} and \IntroPanlSymb{\bfpanel{c}}\IntroPanlSymb{\bfpanel{f}} the data points marked with (red) squares are for the case $\varepsilon=0.25$; with (blue) circles, $\varepsilon=0.5$; and with (green) triangles, $\varepsilon=1$. In plots \IntroPanlSymb{\bfpanel{b}}\IntroPanlSymb{\bfpanel{e}} the (red) squares are also for $\varepsilon=0.25$, while the (green) triangles are for $\varepsilon=2\times 10^{5}$. Plots \bfpanel{a},\bfpanel{b},\bfpanel{c} show the infinite time average, obtained from exact time dynamics, of the ``equipartition measure'' $E_{x}-E_{y}$  versus the value this measure had in the initial state. The straight solid lines are the predictions of equation~(\ref{MAIN_central_result}). Included as insets are the representations of the lattice and its perturbation. In plots \bfpanel{d},\bfpanel{e},\bfpanel{f} we show the level-spacing histograms for the perturbed hamiltonians, which show how close the systems presented in plots \bfpanel{a},\bfpanel{b},\bfpanel{c} are to integrability or to well-developed chaos. The horizontal axis, $\Delta\mbox{e}$, is the level spacing in the unfolded spectrum. The closer the actual distribution to the curve representing the Poissonian level statistics, the more integrable the system; the closer the actual distribution to the curve representing the Gaussian Orthogonal Ensemble (GOE) or {\v S}eba or the Gaussian Unitary Ensemble (GUE), the more chaotic the system. See the main text for details.
}
\end{figure}
%
What we suggest now is to treat the ``real-world'' hamiltonian $\hat{H}$ as a single instance, $(\sigma = \sigma^{\mbox{\scriptsize real-world}},\,\hat{V}=\hat{V}^{\mbox{\scriptsize real-world}})$, of the ensemble of hamiltonians considered above. In this case, the ensemble average on the left-hand side of equation~(\ref{MAIN_almost_the_central_result}) will serve as the ``best predictor'' for the ``real-world'' value of the infinite-time average of the observable, $\inftave{A}$. Notice that, conversely, one of the constituents on the right-hand side of equation~(\ref{MAIN_almost_the_central_result}) will, in its turn, also have to be estimated by a ``best predictor.'' Namely, while $\init{A}$ and $\MC{A}$ are the same for all $(\sigma,\,\hat{V})$-instances and thus can be extracted from a single instance (which can be the``real-world'' instance), the quantity $\etaMC(\sigma,\,\hat{V})$ (on which $\Npca$ depends) varies from instance to instance and so its averaging over instances has a nontrivial effect. Since, however, we only have access to a single realization of $\hat{H}$, we must use the ``real-world'' value of $\etaMC(\sigma,\,\hat{V})$ as the ``best estimate'' for the ensemble average:
$
\Big\langle \etaMC(\sigma,\,\hat{V}) \Big\rangle_{\sigma,\,\hat{V}}
\approx
\etaMC(\sigma^{\mbox{\scriptsize real-world}},\,\hat{V}^{\mbox{\scriptsize real-world}})
=
\etaMC
\,,
$
where
$
\etaMC = \Thrm\left[\etaa\right]
$
is the ``real-world'' microcanonical average of $\etaMC(\sigma,\,\hat{V})$.

An undesirable feature of the expression in equation~(\ref{MAIN_almost_the_central_result}) is that it depends on the number of states in the microcanonical window, $\NinMC$. However, in the limit where $\NinMC$ greatly exceeds the number of principal components,
$
\NinMC \gg \Npca\,,
$
$\NinMC$ disappears from the expression. We finally obtain the desired relationship between the infinite-time average and the initial value of our observable of interest:
\begin{eqnarray}
\inftave{A}
=
\etaMC \init{A}
+
\left(1-\etaMC\right) \MC{A}
\label{MAIN_central_result}
\end{eqnarray}
This is almost our original estimate based on intuitive arguments, except that instead of the typical value of the IPR of the perturbed over the unperturbed states, $\eta_{\alpha}^{\{\vec{n}\}}$, we have the microcanonical average of the IPR of the unperturbed states over the perturbed states. Note the high degree of universality: the same parameter $\etaMC$ is used regardless of what observable $\hat{A}$ one is interested in (which means---since $\hat{A}$ is diagonal in the unperturbed basis---regardless of which integral of motion of $H_{0}(\hat{\vec{n}})$ one is interested in), or with which initial state one starts from among the unperturbed eigenstates.





\paragraph{Testing the analytical prediction}
The relationships in equations~(\ref{MAIN_almost_the_central_result}) and (\ref{MAIN_central_result}) are our central results; we now test them against exact time dynamics of particular physical systems.
Our integrable system will always be a $33 \times 33$-site two-dimensional lattice. We also assume that both $x$- and $y$-cycles of the lattice are threaded by an Aharonov-Bohm (A-B) solenoid each \cite{cheung1988,uski1998,heinrichs2009}. If the lattice is imagined to cover the surface of a torus, the flux is produced by two solenoids: one toroidal, contained within the torus of the lattice, and one straight, passing through the hole of the torus. The corresponding magnetic field fluxes are assumed to be highly irrational but weak (i.e. of the order of one) multiples of the elementary magnetic flux quantum; they are also presumed to be mutually irrational---see the first subsection of the Methods section, below. Their purpose is to destroy the dihedral symmetries of the square lattice while preserving the conservation of momentum in both $x$- and $y$-directions. This lifts the degeneracies in the unperturbed spectrum and randomizes the sequence of appearance of the quantum numbers (two components of the momentum vector) along the energy axis. We investigated three different types of the integrability-breaking perturbation. The first perturbation (see Figure~\ref{f:MAIN_ipr_and_level_statistics}\mfpanel{a}) completes the lattice to a \emphas{single instance} of a Deformed (\emphas{real}) Gaussian Random Matrix Model \cite{kota2001} that couples the eigenstates of the unperturbed lattice. The second perturbation (Figure~\ref{f:MAIN_ipr_and_level_statistics}\mfpanel{b}) is a single impurity of a fixed strength. This system is a lattice version of \v{S}eba billiards \cite{seba1991}, which are known to lie in between integrable and completely quantum-chaotic systems \cite{seba1991}. The third example (Figure~\ref{f:MAIN_ipr_and_level_statistics}\mfpanel{c}) is a \emphas{single instance} of the conventional Anderson disorder \cite{anderson1958}, with a rectangular distribution of the strength in each of the impurities. (We should mention that ergodicity in the context of an Anderson lattice was also studied in ref.~\citen{stotland2008}.) While the first and the second examples correspond to permutation-invariant perturbations, the third one does not. In particular, in comparison to the first two types of perturbation, in the Anderson case there are much stronger correlations between certain kinds of matrix elements. Namely, for each matrix element, consider the momentum difference between the states that the matrix element connects. If one looks at the set of matrix elements which have the same momentum difference, one finds that from one realization of the perturbation to the next, their values all change by the same factor. The observable of interest is the ``equipartition measure''---the difference between the $x$- and $y$-hopping energies, $E_{x}$ and $E_{y}$---whose value in a state of a thermal equilibrium is always zero, thanks to the $x \leftrightarrow y$ symmetry. The infinite-time average of the ``equipartition measure'' $E_{x}-E_{y}$ is shown as a function of its initial value. As the governing parameter $\varepsilon$ we have chosen the ratio between the root mean square of the off-diagonal matrix elements $V_{\vec{n}\vec{n}'}$ (the same for all the pairs $(\vec{n},\,\vec{n}')$) and the typical energy spacing at the energy of interest: $\varepsilon \equiv V_{0} \rho(E)$, where $V_{0} = \sqrt{\overline{|V_{\vec{n}\vec{n}'}|^2}}$ and $\rho(E)$ is the density of states at the energy $E$.

For each type of perturbation, we time-evolved 201 different initial states, taken to be all the eigenstates of the unperturbed lattice whose eigenvalues came from a representative microcanonical energy window; the middle eigenstate (the 101st one in the order of increasing energy within the window) had the energy of $-1.5\,J$, where $J$ is the hopping constant (see the \secMethodsIINumericalIImodelsIIused{} subsection of the Methods section). A \emphas{single instance} of the corresponding hamiltonian (which was random for the cases \bfpanel{a} and \bfpanel{c}) was used in all three cases. The different initial states have different values for $\init{(E_{x}-E_{y})}$, lying between some minimum and maximum values; we needed to group the initial states into sets with similar $\init{(E_{x}-E_{y})}$-values. Thus we partitioned the interval from the minimum to the maximum $\init{(E_{x}-E_{y})}$-value into subintervals centered at $-3 J$, $-2.5 J$, \ldots $3 J$, each of width $\Delta \init{(E_{x}-E_{y})} = .5 J$. All the initial states whose $\init{(E_{x}-E_{y})}$-values fell into the same subinterval constituted a ``group of initial states with similar $\init{(E_{x}-E_{y})}$-values.'' The points shown in plots \bfpanel{a},\bfpanel{b},\bfpanel{c} correspond to the groups: the $x$-value of a point is the center of the subinterval defining the group, while the $y$-value is the group average of $\inftave{(E_{x}-E_{y})}$. The theoretical curves, produced by equation~(\ref{MAIN_central_result}), also involved an averaging: the value of $\init{(E_{x}-E_{y})}$ which is fed into equation~(\ref{MAIN_central_result}) was, for each group, the group average of $\init{(E_{x}-E_{y})}$ rather than the center of the subinterval; this explains why the theoretical curves are not exactly straight lines. As Figures~\ref{f:MAIN_ipr_and_level_statistics}\mfpanel{a},\mfpanel{b},\mfpanel{c} show, the prediction of equation~(\ref{MAIN_central_result}) agrees very well with the numerical results.

Figures~\ref{f:MAIN_ipr_and_level_statistics}\mfpanel{d},\mfpanel{e},\mfpanel{f} are there to show where on the continuum between integrability and well-developed chaos our various systems lie. The discrete points are the plots of the actual level spacing statistics of our systems, properly unfolded \cite{guhr1998}; for the cases \bfpanel{d} and \bfpanel{f}, we also average over 16 realizations of the perturbation $\hat{V}$. For comparison, we also plot the curves corresponding to the completely integrable and (the appropriate) completely quantum-chaotic systems. The curves labeled ``Poisson'' arise for Poisson level statistics, which is characteristic for integrable systems. The most recognizable feature is the nonzero value at zero spacing, representing the absence of level repulsion in integrable systems. The Gaussian Orthogonal Ensemble (GOE) curve is valid for systems with well-developed quantum chaos in the presence of time-reversal invariance; the \v{S}eba distribution \cite{seba1991} holds for systems with singular perturbations; and the Gaussian Unitary Ensemble (GUE) is used in the cases of quantum chaos without time-reversal invariance \cite{guhr1998}. We see that when \mbox{$\varepsilon=0.25$}, the level-spacing statistics of our systems is intermediate between those of the integrable-like and the appropriate completely quantum-chaotic-like distributions. As $\varepsilon$ increases, the level spacing distributions for the real Gaussian, singular, and Anderson perturbations converge, respectively, to the GOE, \v{S}eba, and GUE predictions. (The reason for the GUE statistics in the Anderson case is the Aharonov-Bohm flux, which breaks the time-reversal invariance \cite{uski1998}. In the case of the first model the statistics remains of a GOE type, since the perturbation matrix elements were artificially fixed to real values.) The expression in equation~(\ref{MAIN_central_result}) is thus confirmed in the full range from the integrable regime all the way to the well-developed quantum chaos.

\begin{figure}[h!]
\begin{center}
\includegraphics[scale=.5]{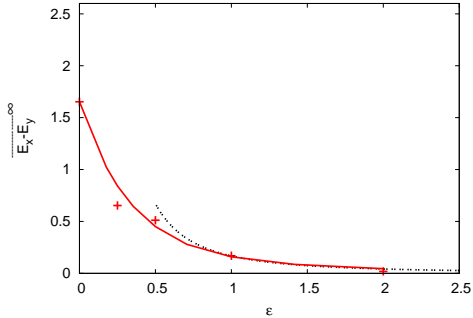}
\end{center}
\caption
{
\label{f:MAIN_kam}
%
%
\textbf{The memory of the initial state as a function of the strength of the integrability-breaking perturbation.}
For the case of Anderson disorder, we compute, for three similar initial states, $|\psi_{\mbox{\scriptsize init.}} \rangle = |n_{x}\!=\!+7,\,n_{y}\!=\!0,\pm 1\rangle$, the infinite time average of the ``equipartition measure'' $E_{x}-E_{y}$  as a function of the disorder strength $\varepsilon$. The red crosses correspond to the averages over the this group of states. The red solid line corresponds to the formula (\ref{MAIN_central_result}) combined with a numerically tabulated values of the inverse participation ratio $\etaMC$ for the case of the Deformed (complex) Gaussian Random Matrix Model. Finally, the black dotted line is given by the asymptotic analytic prediction in equation~(\ref{MAIN_central_result_eps_gg_1}). The lattice parameters are the same as for Figure~\ref{f:MAIN_A_of_time}.
}
\end{figure}

To relate our predictions to the system parameters, we further connect the inverse participation ratio $\etaMC$---otherwise emprirically irrelevant---to the governing parameter $\varepsilon$. This allows us to trace the integrability-to-chaos transition, i.e. express the memory of the initial state through the strength of the non-integrable perturbation. Figure~\ref{f:MAIN_kam} shows, for a particular group of the initial states, the values of the ``equipartition measure'' $\init{(E_{x}-E_{y})}$ as a function of the parameter $\varepsilon$. The numerical values are drawn from the set used to produce Figure~\ref{f:MAIN_ipr_and_level_statistics}. To obtain the theoretical prediction, we tabulate numerically the inverse participation ratio $\eta_{\alpha_{0}}^{\{\alpha\}}$, averaged over the unperturbed states $|\alpha_{0}\rangle$, for a Deformed (complex) Gaussian Random Matrix Model. For large values of $\varepsilon$, we also use known theoretical results \cite{fyodorov1995_R11580,frahm1995_385} for the so-called strength function, $\overline{\left| \langle \alpha_{0} | \alpha \rangle \right|^2}$, combined with the assumption \cite{flambaum1997_5144} of the Gaussian character of the fluctuation of $\langle \alpha_{0} | \alpha \rangle$:
\begin{eqnarray}
\inftave{A} - \MC{A}
\mathrel{ \mathop \approx_{\varepsilon \gtrsim 1}}
\frac{q}{2\pi^2\varepsilon^2} (\init{A}-\MC{A})
\quad,
\label{MAIN_central_result_eps_gg_1}
\end{eqnarray}
where $q=3$ for a non-integrable perturbation that belongs to the Gaussian Orthogonal class, and $q=2$ in the Gaussian
Unitary case. Note that the Aharonov-Bohm flux present in the example of the Figure~\ref{f:MAIN_kam} implies the latter. Note also that the result in equation~(\ref{MAIN_central_result_eps_gg_1}) for the Gaussian Orthogonal case can be confirmed directly \cite{frahm1995_385}, without the assumption of Gaussianity of $\langle \alpha_{0} | \alpha \rangle$. The strength function for an equidistant spectrum perturbed by a real matrix whose matrix elements have the same magnitude and random signs---the case closely related to a Poisson spectrum perturbed by a real Gaussian matrix---first appears in the classic Wigner's paper \cite{wigner1955_548}. The strength function for the latter system {\it per se} is predicted in ref. \citen{jacquod1995_3501}. (See the \secMethodsIIRelevantIItheoreticalIIresults{} subsection of the Methods section for more details.)

\section*{Discussion}
We have thus demonstrated that, in the case when the integrability-breaking perturbation obeys no selection rules and when the spectrum of the underlying integrable system is ``sufficiently irrational,'' it is possible to characterize the memory of the initial values of the unperturbed integrals of motion, as the strength of the perturbation is increased, by a simple and universal expression. The expression was verified for three different types of perturbations away from integrability, including a Deformed Gaussian Random Matrix Model, an isolated impurity, and the two-dimensional Anderson model; in each case, the expression works for a full range of perturbation strengths from completely integrable to completely chaotic.

Our predictions should be testable experimentally with techniques that are already available, or nearly so; likely contexts include those of the investigations of Anderson localization with cold gases in optical lattices \cite{billy2008,roati2008}, or those of quantum mirage configurations of a quantum corral \cite{manoharan2000}, modified to a \v{S}eba-type billiard. In general, memory effects should be visible as soon as the discreteness of energy levels is discernible.

Future work must try to come to terms with cases where the partial preservation of the integrals of motion is enhanced via some selection rules obeyed by the perturbing potential. The two most empirically relevant examples of selection rules come from the limitations imposed by the few-site nature of the hopping terms in typical lattice hamiltonians, and by the few-body nature of interactions in many-body systems. Encouragingly, some work relevant to these cases has already been done; ref.~\citen{brown2010} studied the role of the former type of selection rules, while the effects of the latter type on the structure of the eigenstates has been investigated in a number of works; for a review, see ref.~\citen{kota2001} and the references therein, particularly ref.~\citen{flambaum1996_5729}. More generally, one should start systematically increasing the complexity of the topology of the network of transitions.

\section*{Methods}
%
%


\paragraph{Numerical models used to verify equation~(\ref{MAIN_central_result}).}

As an example of an unperturbed hamiltonian $\hat{H}_{0}$, we use a $N_{x} \times N_{y}$ ($33 \times 33$ in the numerical examples considered) two-dimensional lattice with periodic boundary conditions and odd $N_{x}$, $N_{y}$:
\begin{eqnarray}
\hat{H}_{0}
&=&
 -J \!\!\sum_{j_{x}=-\frac{N_{x}-1}{2}}^{+\frac{N_{x}-1}{2}} \sum_{j_{y}=-\frac{N_{y}-1}{2}}^{+\frac{N_{y}-1}{2}}
\!
                                           \left(
                                                e^{-i \frac{2\pi \Delta n_{x}}{N_{x}} } |(j_{x}+1,\,j_{y})\rangle \langle (j_{x},\,j_{y})|
                                                +
                                           \right.
\notag
\\
&& \qquad
                                           \left.
                                                e^{-i \frac{2\pi \Delta n_{y}}{N_{y}} } |(j_{x},\,j_{y}+1)\rangle \langle (j_{x},\,j_{y})|
                                                +
                                                h.c.
                                           \right) ,
\label{MAIN_Integrable_hamiltonian}
\end{eqnarray}
where $|(j_{x},\,j_{y})\rangle$ are the eigenstates of position. Both $x$- and $y$-hopping constants have the same amplitude $J$. We also assume that both $x$- and $y$-cycles of the lattice are threaded by an Aharonov-Bohm (A-B) solenoid each. This leads to the hopping constants acquiring complex phase factors, with phases $-2\pi\Delta n_{x}/N_{x}$ and $-2\pi\Delta n_{y}/N_{y}$ respectively. Here $\Delta n_{x} \equiv \phi_{x}/\phi_{0}$ ($\Delta n_{y} \equiv \phi_{y}/\phi_{0}$) where $\phi_{x}$ ($\phi_{y}$)
is the magnetic flux through the $x$-cycle ($y$-cycle), $\phi_{0} = 2\pi \hbar c/q$ is the elementary quantum of magnetic flux, $c$ is the speed of light, and $q$ is the electric charge of the lattice particle. The flux is assumed to be very weak, $\Delta n_{x}(\Delta n_{y}) \sim 1$. Note also that after a suitable gauge transformation, the complex hoppings can be replaced by real ones, supplemented by twisted boundary conditions \cite{cheung1988,uski1998,heinrichs2009}. The eigenstates of the unperturbed hamiltonian are plane waves $|(n_{x},\,n_{y})\rangle$,
\begin{eqnarray*}
\langle (j_{x},\,j_{y})|(n_{x},\,n_{y})\rangle
=(N_{x} N_{y})^{-1/2}\exp\left(i \frac{2\pi n_{x}j_{x}}{N_{x}}+i \frac{2\pi n_{y}j_{y}}{N_{y}} \right)\,.
\end{eqnarray*}
The linear momentum quantum numbers $n_{x}$ and $n_{y}$ ($|n_{x}|\leq(N_{x}-1)/2$ and $|n_{y}|\leq(N_{y}-1)/2$) constitute a set of the integrals of motion. Eigenenergies of the unperturbed hamiltonian are
\begin{eqnarray*}
E_{n_{x},\,n_{y}}=-2J \left\{\cos[2\pi(n_{x}+\Delta n_{x})/N_{x}] +
\cos[2\pi(n_{y}+\Delta n_{y})/N_{y}]\right\}.
\end{eqnarray*}

The purpose of introducing the A-B flux is to generate an equi-energy surface $E_{n_{x},\,n_{y}} = E$ that is sufficiently irrational with respect to the lattice of the integer quantum numbers $n_{x}$ and $n_{y}$. To this end, the ``defects'' $\Delta n_{x}$ and $\Delta n_{y}$ must be both irrational and mutually irrational (see the first subsection of the first section of the Supplementary Methods and Supplementary Figure~\SIfIISUPPIIHzeroIIrandomization{}). In our numerical calculations, we used $\Delta n_{x}  = \varphi/4$ and $\Delta n_{y} =  e/10$, where  $\varphi = 1.618\ldots$ is the golden ratio and $e =2.718\ldots $ is the base of the natural logarithm.

As the first example of an integrability-breaking perturbation $\hat{V}$, we considered a member of a Gaussian Orthogonal Ensemble acting between the eigenstates of the unperturbed lattice (Figure~{\ref{f:MAIN_ipr_and_level_statistics}a):
\begin{eqnarray*}
V_{\vec{n}\vec{n}'} = \sqrt{1+\delta_{\vec{n}\vec{n}'}} V_{0} \xi_{\vec{n}\vec{n}'}
\quad,
\end{eqnarray*}
where the $\xi_{\vec{n}\vec{n}'}$ are $N_{x}N_{y} (N_{x}N_{y} + 1)/2$ independent real Gaussian-distributed random variables of unit variance and zero mean, with the remaining $N_{x}N_{y} (N_{x}N_{y} - 1)/2$ matrix elements controlling the hermiticity of $\hat{V}$.

The second example (Figure~{\ref{f:MAIN_ipr_and_level_statistics}b) is a singular perturbation,
\begin{eqnarray*}
V(j_{x},\,j_{y}) = V_{0} N_{x} N_{y} \delta_{(j_{x},\,j_{y}),\,(0,\,0)}
\quad,
\end{eqnarray*}
which produces a matrix with all-equal matrix elements:
\begin{eqnarray*}
V_{\vec{n}\vec{n}'} = V_{0}
\quad.
\end{eqnarray*}
In this case, the eigenstates of the perturbed hamiltonian can be found exactly. They read
\begin{eqnarray*}
|\alpha\rangle=C_{\alpha}^\text{sing}
\sum_{n_{x}=-\frac{N_{x}-1}{2}}^{+\frac{N_{x}-1}{2}} \sum_{n_{y}=-\frac{N_{y}-1}{2}}^{+\frac{N_{y}-1}{2}}
\frac{|(n_{x},\,n_{y})\rangle}{E_{\alpha}-E_{n_{x},\,n_{y}}}
\quad,
\end{eqnarray*}
where the corresponding eigenenergies $E_{\alpha}$ are the solutions of the algebraic equation
\begin{eqnarray*}
\sum_{n_{x}=-\frac{N_{x}-1}{2}}^{+\frac{N_{x}-1}{2}} \sum_{n_{y}=-\frac{N_{y}-1}{2}}^{+\frac{N_{y}-1}{2}}
\frac{1}{E_{\alpha}-E_{n_{x},\,n_{y}}}=\frac{1}{V_{0}}
\quad,
\end{eqnarray*}
and the normalization factor is defined by
\begin{eqnarray*}
\left(C_{\alpha}^\text{sing}\right)^{-2}=
\sum_{n_{x}=-\frac{N_{x}-1}{2}}^{+\frac{N_{x}-1}{2}} \sum_{n_{y}=-\frac{N_{y}-1}{2}}^{+\frac{N_{y}-1}{2}}
\left(E_{\alpha}-E_{n_{x},\,n_{y}}\right)^{-2}
\quad.
\end{eqnarray*}
A solution of this form was at first obtained by \v{S}eba \cite{seba1990}, for a flat continuous billiard. A similar problem involving a two-dimensional lattice with periodic boundary conditions in one direction and a trapping potential in another was recently analyzed \cite{valiente2011}.

Finally, we consider an Anderson-type disorder (Figure~{\ref{f:MAIN_ipr_and_level_statistics}c):
\begin{eqnarray*}
V(j_{x},\,j_{y}) = W \zeta_{j_{x},\,j_{y}}
\quad,
\end{eqnarray*}
where
the
$\zeta_{j_{x},\,j_{y}}$ are $N_{x} N_{y}$ real independent variables, distributed uniformly between $-1$ and $+1$; $W$ is the Anderson disorder parameter. The interaction strength parameter $V_{0}$ used in the previous cases corresponds to the r.m.s.\ of the (generally complex) off-diagonal matrix elements $V_{\vec{n}\vec{n}'}$, $V_{0} \equiv \sqrt{\overline{|V_{\vec{n}\vec{n}'}|^2}} = W/(2 \sqrt{3} \sqrt{N_{x} N_{y}})$.

%
\paragraph{Relevant results on the IPR in random matrix models.}

To obtain the solid line in Figure~\ref{f:MAIN_kam}, we computed the inverse participation ratio $\eta_{\alpha_{0}}^{\{\alpha\}}$, averaged over all unperturbed states $| \alpha_{0} \rangle$, for a $N \times N$ Deformed Gaussian Random Matrix $\hat{h} = \hat{h}_{0} + \hat{v}$, with $N=2000$. The ``integrable'' part of the matrix, $\hat{h}_{0}$, was represented by a diagonal matrix whose $N$ diagonal entries were given by $N$ independent real random numbers uniformly distributed in the interval $[-(N/\rho)/2,\, +(N/\rho)/2]$. Here, $\rho$ is the density of states. The ``non-integrable part'', $\hat{v}$, was a random matrix drawn from the Gaussian Unitary Ensemble of random Hermitian matrices: (real) diagonal matrix elements and the real and imaginary parts of the off-diagonal matrix elements above the diagonal (those below the diagonal then being fixed by hermiticity) were given by independent, Gaussian-distributed random numbers with standard deviations $\sigma_{\mbox{\scriptsize diag.}} = V_{0}$ and $\sigma_{\mbox{\scriptsize off-diag., Re}} = \sigma_{\mbox{\scriptsize off-diag., Im}} = V_{0}/\sqrt{2}$, respectively. For completeness, the Gaussian Orthogonal case would give $\sigma_{\mbox{\scriptsize diag.}} = \sqrt{2} V_{0}$ and $\sigma_{\mbox{\scriptsize off-diag.}} = V_{0}$. In both cases, $V_{0}$ fixes the mean square of the off-diagonal matrix elements: $\overline{|v_{\alpha_{0}^{},\alpha_{0}'}|^2} = V_{0}^2$.

Equation~(\ref{MAIN_central_result_eps_gg_1}) uses the following asymptotic expression for the inverse participation ratio $\eta$:
\begin{eqnarray}
\eta
\mathrel{ \mathop \approx_{\varepsilon \gtrsim 1}}
\frac{q}{2\pi^2\varepsilon^2}
\quad,
\label{eta_eps_gg_1}
\end{eqnarray}
where $q=3$ for a non-integrable perturbation that belongs to the Gaussian Orthogonal class, and $q=2$ in the Gaussian Unitary case. In refs. \citen{fyodorov1995_R11580,frahm1995_385,wigner1955_548} and \citen{jacquod1995_3501} it was shown that for $\varepsilon \gg 1$, the strength function converges to
\begin{eqnarray}
\overline{\left| \langle \alpha_{0} | \alpha \rangle \right|^2}
\mathrel{ \mathop \approx_{\varepsilon \gtrsim 1}}
\frac{1}{\pi\rho} \frac{\hbar\Gamma/2}{    (E_{\alpha_{0}}-E_{\alpha})^2 + (\hbar\Gamma/2)^2 }
\quad,
\label{BW}
\end{eqnarray}
in both Gaussian Orthogonal and Gaussian Unitary cases. Here, $\Gamma $ is determined by the Fermi Golden Rule: $\Gamma = 2\pi V_{0}^2 \rho /\hbar$.
Also, it is known \cite{flambaum1997_5144} that in the same limit, the individual coefficients $\langle \alpha_{0} | \alpha \rangle$ behave as independent Gaussian random variables, with zero mean and with a standard deviation governed by the expression (\ref{BW}). Recall that the fourth moment of a Gaussian distribution of a variable $\xi$ is related to the second one as \cite{Sayed2008} $\overline{|\xi|^4} = q \left(\overline{|\xi|^2}\right)^2$. Also, it is important that when $\varepsilon \gg 1$, the energy width $\Gamma$ in equation~(\ref{BW}) contains many levels, and thus the sum in the definition of the inverse participation ratio, $\eta_{\alpha_{0}}^{\{\alpha\}} \equiv \sum_{\alpha} |\langle \alpha | \alpha_{0} \rangle |^4$, can be replaced by an integral. The asymptotic formula (\ref{eta_eps_gg_1}) for the inverse participation ratio immediately follows.

\paragraph{\textit{Ab initio} computations.} All numerical time evolution was computed from full exact diagonalization of the hamiltonians using the \textit{Mathematica} computer package.

\begin{addendum}
\item[\textsf{Acknowledgements}] \mbox{}\\We are grateful to F. Werner and D. Cohen for enlightening discussions on the subject. 
Supported by the Office of Naval Research grants N00014-09-1-0502 (M.O. and
V.D.) and N00014-09-1-0966 (M.R.), and the National Science Foundation grants PHY-1019197 (M.O. and V.D.) and PHY-0902906 (K.J.).

\end{addendum}

\renewcommand{\figurename}{Supplementary Figure}
\renewcommand{\thefigure}{S\arabic{figure}}
\renewcommand{\thetable}{S\arabic{table}}
\renewcommand{\theequation}{S-\arabic{equation}}
\renewcommand{\thesection}{}
\renewcommand{\thesubsection}{SD\arabic{subsection}}
\renewcommand{\refname}{Supplementary References}

\newcommand{\MAINIIalmostIItheIIcentralIIresult}{2}
\newcommand{\MAINIIcentralIIresult}{3}
\newcommand{\MAINIIcentralIIresultIIepsIIggIIone}{4}
\newcommand{\MAINIIIntegrableIIhamiltonian}{5}

\newcommand{\fIIMAINIIiprIIandIIlevelIIstatistics}{2}
\newcommand{\fIIMAINIIkam}{3}







\newcommand{\secOne}{}
\newcommand{\secTwo}{}
\newcommand{\secThree}{}
\newcommand{\secFour}{}
\newcommand{\secFive}{}

\renewcommand{\secMethodsIINumericalIImodelsIIused}[1]{first}
\renewcommand{\secMethodsIIRelevantIItheoreticalIIresults}[1]{second}

\newcommand{\citemain}[1]{$^{#1}$}
\newcommand{\citenmain}[1]{#1}
\newcommand{\refkota}{\! 19}
%
%
\newpage
\section{\mbox{}\hspace{-1em}Supplementary Figures}

\begin{figure}[ht!]
\mbox{}\vspace{\baselineskip}
\begin{center}
\includegraphics[scale=.26]{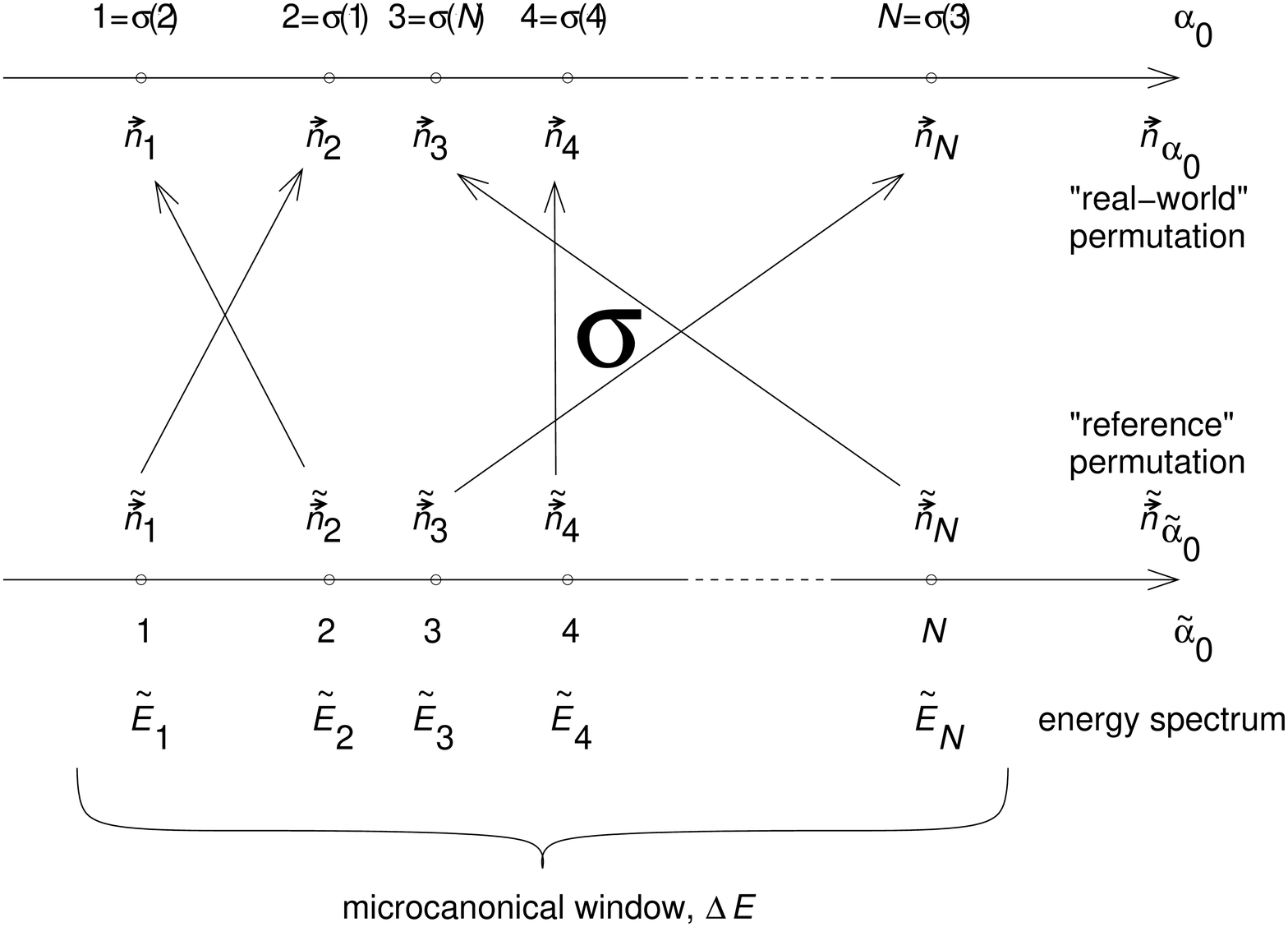}
\end{center}
\caption
{
\label{f:SUPP_ho_randomization_illustration}
\textbf{The energy-ordered sequence of the eigenstates of a ``real-world'' integrable system viewed as a random permutation of an analogous sequence of the eigenstates of a fictitious ``reference'' hamiltonian.}
}
\end{figure}


\newpage
\begin{figure}[ht!]
\begin{center}
\includegraphics[scale=.37]{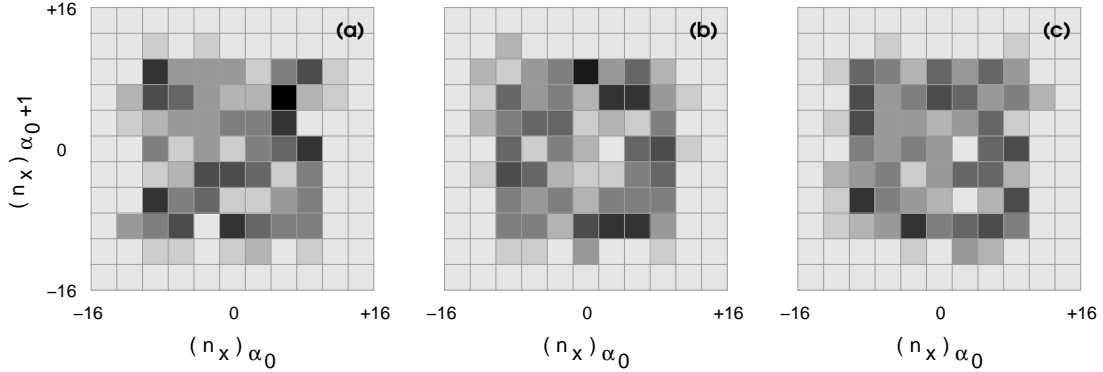}
\end{center}
\caption
{
\label{f:SUPP_H0_randomization}
\textbf{A test verifying that the integrable hamiltonian in equation~(\MAINIIIntegrableIIhamiltonian) of the main text has a ``sufficiently irrational'' equi-energy surface. } We find no correlations between the quantum numbers of pairs of eigenstates that are nearest neighbors in a sequence ordered by increasing energy: \IntroPanlSymb{\bfpanel{a}} A two-dimensional histogram for the distribution of the $\left( (n_{x})_{\alpha_{0}},\,(n_{x})_{\alpha_{0}+1} \right)$ pairs of the values of the $x$-momentum of two consecutive eigenstates (the index $\alpha_{0}$ labels the eigenstates in order of increasing energy), showing no apparent correlation between $(n_{x})_{\alpha_{0}+1}$ and $ (n_{x})_{\alpha_{0}}$. \IntroPanlSymb{\bfpanel{b}}\IntroPanlSymb{\bfpanel{c}} The same histogram as in plot \bfpanel{a} for two fictitious sets of eigenstates obtained via two random permutations of the original sequence (see also Supplementary Figure~\ref{f:SUPP_ho_randomization_illustration}), presented for comparison with the histogram in plot \bfpanel{a}: the latter, which is the one for the actual system, looks as random as the histograms in plots \bfpanel{b} and \bfpanel{c}, which were generated by random permutations of eigenstates. The integrable hamiltonian was a $33 \times 33$-site lattice pierced by a weak Aharonov-Bohm (A-B) flux. The histograms include states from a representative microcanonical energy window comprising $201$ eigenstates with the energy of the middle ($101$st) state being $E=-1.5\,J$, where $J$ is the hopping constant. Individual boxes of the histogram contain $3 \times 3 = 9$ values of $\left( (n_{x})_{\alpha_{0}},\,(n_{x})_{\alpha_{0}+1} \right)$. For other details, see the beginning of the \secMethodsIINumericalIImodelsIIused{} subsection of the Methods section of the main text.
}
\end{figure}

%
\newpage
\section{\mbox{}\hspace{-1em}Supplementary Methods}
\subsection*{A model supporting the formula in equation~({\MAINIIalmostIItheIIcentralIIresult}) of the main text}
%
\subsubsection{The unperturbed integrable hamiltonian.}
Consider the hamiltonian of an integrable quantum system under a nonintegrable perturbation $\hat{V}$:
\begin{eqnarray*}
\hat{H} = H_{0}(\hat{\vec{n}}) + \hat{V}
\quad,
\end{eqnarray*}
where $\hat{\vec{n}} = \left\{\hat{n}_{1},\,\hat{n}_{2},\,\ldots,\,\hat{n}_{d} \right\}$ is a complete set of the integrals of motion of the hamiltonian $H_{0}(\hat{\vec{n}})$, and $d$ is the number of the degrees of freedom. If the system is large enough, one should be able to identify a \textit{macroscopic energy scale} ${\cal E}$ that governs the regular (as opposed to fluctuating) part of the dependence of thermal expectation values of observables on energy. Under these circumstances, it is possible to choose, for a given energy $E$, a \textit{microcanonical window}
\begin{eqnarray}
{\cal W}_{\mbox{\scriptsize MC}}(E,\,\Delta E) = [E-\Delta E/2,\,E+\Delta E/2]
\label{MC_window}
\end{eqnarray}
that is small compared to the macroscopic scale ${\cal E}$ but still large enough to contain a large number of the eigenstates. Let us arrange the eigenstates of $H_{0}(\hat{\vec{n}})$ inside the window in the order of increasing energy and label them using an index $\alpha_{0} = 1,\,\ldots,\,N$:
\begin{eqnarray}
&&
\alpha_{0} > \alpha_{0}' \, \Rightarrow \, E_{\vec{n}_{\alpha_{0}^{}}} > E_{\vec{n}_{\alpha_{0}'}}
\label{SUPP_ordering}
\\
&&
E_{\vec{n}_{\alpha_{0}=1}} \approx E-\Delta E/2;  \quad E_{\vec{n}_{\alpha_{0}=N}} \approx E+\Delta E/2
\nonumber
\quad,
\end{eqnarray}
where $N$ is the number of the eigenstates the window contains.
The hamiltonian of the unperturbed system reads
\begin{eqnarray}
\hat{H}_{0} = \sum_{\alpha_{0} = 1}^{N} \tilde{E}_{\alpha_{0}} |\vec{n}_{\alpha_{0}} \rangle \langle \vec{n}_{\alpha_{0}} |
+
\ldots
\label{SUPP_H_0}
\quad,
\end{eqnarray}
where $|\vec{n}\rangle$ and $E_{\vec{n}}$ are the eigenstates and eigenenergies respectively,
\begin{eqnarray*}
\tilde{E}_{\alpha_{0}} \equiv E_{\vec{n}_{\alpha_{0}}}
\quad,
\end{eqnarray*}
and ``$\ldots$'' stands for the states outside the microcanonical window.

Now, we suggest replacing the hamiltonian in equation~(\ref{SUPP_H_0}) by an ensemble of random hamiltonians:
\begin{eqnarray}
&&
\hat{H}_{0}(\sigma) = \sum_{\alpha_{0} = 1}^{N} \tilde{E}_{\alpha_{0}}
|\vec{n}_{\alpha_{0}}(\sigma) \rangle
\langle \vec{n}_{\alpha_{0}}(\sigma) |
+
\ldots
\label{SUPP_ensemble_H_0}
\\
&&
\vec{n}_{\alpha_{0}}(\sigma) \equiv \tilde{\vec{n}}_{\tilde{\alpha}_{0}=\sigma^{-1}(\alpha_{0})}
\quad,
\nonumber
\end{eqnarray}
where
$\tilde{\vec{n}}_{\tilde{\alpha}_{0}}$ is a ``reference'' sequence of the eigenstates, and the permutations $\sigma$ are assumed to occur with equal probability $p=1/N!$\,\,. The ``real-world'' sequence $\vec{n}_{\alpha_{0}}$ is supposed to be represented by a particular realization of a random hamiltonian from equation~(\ref{SUPP_ensemble_H_0}) corresponding to a particular permutation $\sigma^{{\mbox{\scriptsize real-world}}}$:
\begin{eqnarray*}
\vec{n}_{\alpha_{0}} = \vec{n}_{\alpha_{0}}(\sigma^{\mbox{\scriptsize real-world}})
\quad.
\end{eqnarray*}
Notice that the spectra and the sets of the eigenstates are identical for all members of the ensemble; however, the order in which the eigenstates appear on the energy axis is different, and is realized randomly. The relationship between the reference sequence of eigenstates and the sequence of eigenstates as it is for the actual, ``real-world''  integrable hamiltonian is illustrated in Supplementary Figure~\ref{f:SUPP_ho_randomization_illustration}.


The replacement of the actual integrable hamiltonian by an ensemble of random integrable hamiltonians is justified if the sequence of eigenstates in the actual integrable hamiltonian is indeed ``sufficiently random.'' In the main text, this property was phrased in terms of the integrable hamiltonian having a ``sufficiently irrational'' equi-energy surface: let $\{(\hat{n}_{k})_{\indx}\}$ be the sequence (labeled by $\indx$) of the values of the $k$th quantum number as one is going from one eigenstate of $H_{0}(\vec{n})$ to the next in the order of increasing energy. The equi-energy surface is said to be sufficiently irrational if, for every $k$, the sequence $\{(\hat{n}_{k})_{\indx}\}$ passes any simple statistical test for randomness. One such test is explained in Supplementary Figure~\ref{f:SUPP_H0_randomization} on the example of the integrable hamiltonian defined in detail in the \secMethodsIINumericalIImodelsIIused{} subsection of the Methods section of the main text.
\subsubsection{The non-integrable perturbation.}
Let us first assume that the perturbation $\hat{V}$ is weak as compared to the width of the microcanonical window,
\begin{eqnarray*}
|V_{\vec{n}\vec{n}'}| \ll \Delta E
\quad,
\end{eqnarray*}
so that it mostly couples the states inside the microcanonical window of equation~(\ref{MC_window}) to themselves. In this case we can approximate the perturbation by its truncated version, i.e.\ we neglect all the matrix elements that involve the states outside the window in equation~(\ref{MC_window}):
\begin{eqnarray*}
V_{\vec{n}\vec{n}'} \neq 0 \, \Rightarrow \, E_{\vec{n}} \in {\cal W}_{\mbox{\scriptsize MC}}(E,\,\Delta E) \,\, \& \, \, E_{\vec{n}'} \in {\cal W}_{\mbox{\scriptsize MC}}(E,\,\Delta E)
\quad,
\end{eqnarray*}

Let's further assume that the perturbation $\hat{V}$ does not obey any apparent selection rules. In this case, $\hat{V}$ can be thought as a particular realization of a random matrix, with the distribution of the matrix elements, $w(\hat{V})$, invariant under the permutations of the eigenstates:
\begin{eqnarray}
w(\hat{U}_{\sigma}\hat{V}\hat{U}_{\sigma}^{-1}) = w(\hat{V})
\quad,
\label{V_invariance}
\end{eqnarray}
where
\begin{eqnarray*}
\hat{U}_{\sigma} | \vec{n}_{\alpha_{0}} \rangle = | \vec{n}_{\sigma(\alpha_{0})} \rangle
\quad.
\end{eqnarray*}
The Gaussian Orthogonal (respectively, Unitary) ensemble is the most obvious candidate for such a distribution: its distribution of the matrix elements (real for the Gaussian Orthogonal ensemble and complex for the Gaussian Unitary),
\begin{eqnarray*}
w_{\mbox{\scriptsize GOE(GUE)}}(\hat{V}) \propto \exp[-\mbox{Tr}(\hat{V}^2)/4 V_{0}^2]
\quad,
\end{eqnarray*}
is invariant under orthogonal (respectively, unitary) transformations, and in particular under permutations. Here, $V_{0} = \sqrt{\overline{|V_{\vec{n}\vec{n}'}|^2}}$ is the standard deviation of the off-diagonal elements. A (fixed, non-fluctuating) singular perturbation,
\begin{eqnarray*}
w_{\mbox{\scriptsize singular}}(\hat{V}) \propto \prod_{\vec{n}} \prod_{\vec{n}'} \delta(V_{\vec{n}\vec{n}'} - V_{0})
\quad,
\end{eqnarray*}
is another example of a ``permutation-invariant'' distribution of the perturbation matrices.



\subsection*{A proof of the formula in equation~({\MAINIIalmostIItheIIcentralIIresult}) in the main text}
%
For a given permutation $\sigma$ and a given set of the matrix elements of the perturbation $\hat{V}$, the infinite time average of the quantum expectation value of an observable $\hat{A}$ will be given by
\begin{eqnarray}
\inftave{A}(\sigma,\,\hat{V})
&\equiv&
 \lim_{\tmax\to\infty} \frac{1}{\tmax} \int_{0}^{\tmax} \! dt \, \langle \psi(t) | \hat{A} | \psi(t) \rangle
\nonumber
\\
&=&
\sum_{\alpha} \left| \langle \alpha | \vec{n}_{\mbox{\scriptsize init.}} \rangle \right|^2 \langle \alpha | \hat{A} | \alpha \rangle
\nonumber
\\
&=&
\eta^{\{\alpha\}}_{\vec{n}_{\mbox{\scriptsize init.}}}\!(\sigma,\,\hat{V}) \, \init{A}
\label{SUPP_A_relax_1}
+
\sum_{\vec{n} \neq \vec{n}_{\mbox{\scriptsize init.}}} F_{\vec{n}\vec{n}_{\mbox{\scriptsize init.}}}\!(\sigma,\,\hat{V}) \, A_{\vec{n}}
\,\,,
%
\end{eqnarray}
The significance of singling-out the first term in the last line will become apparent later in the derivation (see the remark after equation~(\ref{SUPP_A_relax_4})). Here
\begin{eqnarray*}
\init{A} \equiv A_{\vec{n}_{\mbox{\scriptsize init.}}}
\end{eqnarray*}
is the initial quantum expectation value of the observable; $|\alpha \rangle$ is an eigenstate of the perturbed hamiltonian, of an eigenenergy $E_{\alpha}$:
\begin{eqnarray*}
\hat{H} |\alpha\rangle = E_{\alpha} |\alpha\rangle
\quad;
\end{eqnarray*}
the quantity
\begin{eqnarray}
\etaa(\sigma,\,\hat{V}) \equiv
\sum_{\alpha =1}^{N} |\langle \alpha | \vec{n} \rangle|^4
\label{SUPP_IPR}
\end{eqnarray}
is the \textit{inverse participation ratio} of an unperturbed eigenstate $| \vec{n} \rangle$ over the perturbed eigenstates $|\alpha\rangle$; the function
\begin{eqnarray}
F^{\{\alpha\}}_{\vec{n}\vec{n}'}(\sigma,\,\hat{V}) \equiv
\sum_{\alpha =1}^{N} |\langle \alpha | \vec{n} \rangle|^2 |\langle \alpha | \vec{n}' \rangle|^2
\label{SUPP_F}
\end{eqnarray}
is the so-called $F$-function\citemain{\refkota}. Here and below, we assume that the initial state, $|\psi(t\!=\!0)\rangle = | \vec{n}_{\mbox{\scriptsize init.}} \rangle$, is one of the eigenstates of the unperturbed hamiltonian from equation~(\ref{SUPP_ensemble_H_0}) that belong to the window of equation~(\ref{MC_window}), and that the observable $\hat{A}$ is diagonal in the eigenbasis of the unperturbed hamiltonian: $\langle \vec{n} | \hat{A} | \vec{n}' \rangle = A_{\vec{n}} \delta_{\vec{n}\,\vec{n}'}$. As it was mentioned above, we neglect all matrix elements of $\hat{V}$ that couple the inside of the window with the outside.

One can replace the $\vec{n}$ indices in equations~(\ref{SUPP_IPR}-\ref{SUPP_F}) by the $\alpha_{0}$ indices:
\begin{eqnarray}
&&
\eta^{\{\alpha\}}_{\alpha_{0}}(\sigma,\,\hat{V})
\equiv
\eta^{\{\alpha\}}_{\vec{n}_{\alpha_{0}}(\sigma)}(\sigma,\,\hat{V})
\\
&&
F^{\{\alpha\}}_{\alpha_{0}^{}\alpha_{0}'}(\sigma,\,\hat{V})
\equiv
F^{\{\alpha\}}_{\vec{n}_{\alpha_{0}^{}}(\sigma)\vec{n}_{\alpha_{0}'}(\sigma)}(\sigma,\,\hat{V})
\quad,
\end{eqnarray}
see equation~(\ref{SUPP_ordering}). The infinite-time average of the observable now becomes
\begin{eqnarray*}
\inftave{A}(\sigma,\,\hat{V})
&=&
\eta^{\{\alpha\}}_{\alpha_{0,\mbox{\scriptsize init.}}(\sigma)}\!(\sigma,\,\hat{V}) \, \init{A}
\\
&&\,
+
\!\!
\sum_{\alpha_{0} \neq \alpha_{0,\mbox{\scriptsize init.}}(\sigma)} \!\! F^{\{\alpha\}}_{\alpha_{0}\alpha_{0,\mbox{\scriptsize init.}}(\sigma)}\!(\sigma,\,\hat{V})
\, A_{\vec{n}_{\alpha_{0}^{}}(\sigma)}
\,\,,
\end{eqnarray*}
where $\alpha_{0,\mbox{\scriptsize init.}}(\sigma) \equiv \sigma(\tilde{\alpha}_{0,\mbox{\scriptsize init.}})$ is the position of the initial state in the sequence given by a permutation $\sigma$, and $\tilde{\alpha}_{0,\mbox{\scriptsize init.}}$ is the position of the initial state in the ``reference'' sequence: $\tilde{\vec{n}}_{\tilde{\alpha}_{0,\mbox{\scriptsize init.}}}=\vec{n}_{\mbox{\scriptsize init.}}$.

Next, consider the average of the infinite-time average of the observable over the realizations of the perturbation $\hat{V}$:
\begin{eqnarray}
\Big\langle \inftave{A} \Big\rangle_{\hat{V}} (\sigma)
&=&
\Big\langle \eta^{\{\alpha\}}_{\alpha_{0,\mbox{\scriptsize init.}}(\sigma)} \Big\rangle_{\hat{V}} \, \init{A}
\label{SUPP_A_relax_3}
\\
&&\!\!\!
+
\!\!
\sum_{\alpha_{0} \neq \alpha_{0,\mbox{\scriptsize init.}}(\sigma)} \!\!
\Big\langle F^{\{\alpha\}}_{\alpha_{0}\alpha_{0,\mbox{\scriptsize init.}}(\sigma)} \Big\rangle_{\hat{V}} \, A_{\vec{n}_{\alpha_{0}^{}}(\sigma)}
\quad.
\nonumber
\end{eqnarray}
Notice that we used the property that, by construction, the IPR and the $F$-function, if expressed as functions of $\alpha_{0}$, depend on the permutation $\sigma $ only through the perturbation matrix elements $V_{\alpha_{0}^{}\alpha_{0}'} \equiv V_{\vec{n}_{\alpha_{0}^{}}(\sigma)\, \vec{n}_{\alpha_{0}'}(\sigma)}$. Therefore, by virtue of the invariance of the distribution of $\hat{V}$ with respect to the permutations, (see (\ref{V_invariance})), the average IPR and the $F$-function (again, expressed through the $\alpha_{0}$ indices) do not depend on the permutation. That is, for any two permutations $\sigma$ and $\sigma'$,
\begin{eqnarray}
&&
\Big\langle \eta^{\{\alpha\}}_{\alpha_{0}} \Big\rangle_{\hat{V}}(\sigma)
=
\Big\langle \eta^{\{\alpha\}}_{\alpha_{0}} \Big\rangle_{\hat{V}}(\sigma')
=
\Big\langle \eta^{\{\alpha\}}_{\alpha_{0}} \Big\rangle_{\hat{V}}
\nonumber
\\
&&
\Big\langle F^{\{\alpha\}}_{\alpha_{0}^{}\alpha_{0}'} \Big\rangle_{\hat{V}}(\sigma)
=
\Big\langle F^{\{\alpha\}}_{\alpha_{0}^{}\alpha_{0}'} \Big\rangle_{\hat{V}}(\sigma')
=
\Big\langle F^{\{\alpha\}}_{\alpha_{0}^{}\alpha_{0}'} \Big\rangle_{\hat{V}}
\quad.
\label{F_invariance}
\end{eqnarray}

Now, let us consider the average $\Big\langle \inftave{A} \Big\rangle_{\hat{V},\sigma}$, which is  the average of the infinite-time average of the observable over both the ensemble of $\hat{V}$'s and the ensemble of the unperturbed hamiltonians whose members are parametrized, as before, by the permutations $\sigma$. Since $\init{A}$ is invariant over permutations $\sigma$, the $\sigma$-average of the first term in the r.h.s.\ of equation~(\ref{SUPP_A_relax_3}) becomes
\begin{eqnarray}
\Big\langle \Big\langle \eta^{\{\alpha\}}_{\alpha_{0,\mbox{\scriptsize init.}}(\sigma)} \Big\rangle_{\hat{V}} \, \init{A} \Big\rangle_{\sigma}
=
\Big\langle \etaMC(\sigma,\,\hat{V}) \Big\rangle_{\sigma,\,\hat{V}} \, \init{A}
\,,
\label{SUPP_A_relax_3.5}
\end{eqnarray}
where
\begin{eqnarray}
\etaMC(\sigma,\,\hat{V})
&\equiv&
N^{-1} \sum_{\alpha_{0}=1}^{N}  \eta^{\{\alpha\}}_{\alpha_{0}}(\sigma,\,\hat{V})
\label{SUPP_eta_therm_1}
\\
&=&
N^{-1} \!\! \sum_{\vec{n}:\,E_{\vec{n}} \in {\cal W}_{\mbox{\scriptsize MC}}(E,\,\Delta E)}^{N} \, \etaa(\sigma,\,\hat{V})
\nonumber
\end{eqnarray}
is the \textit{microcanonical average of the inverse participation ratio}, for a hamiltonian parametrized by a permutation $\sigma$ and the perturbation matrix $\hat{V}$. Note that formally speaking, the $\sigma$-averaging in the r.h.s.\ of equation~(\ref{SUPP_A_relax_3.5}) is not necessary. However, in any approximate $\hat{V}$-average over a finite series, the $\sigma$-dependence of $\Big\langle \etaMC(\sigma,\,\hat{V}) \Big\rangle_{\hat{V}}$ will inevitably appear again.

The $\sigma$-averaging of the second term requires a two-stage procedure. First, we will average only over the permutations for which the initial state index $\alpha_{0,\mbox{\scriptsize init.}}(\sigma)$ is fixed to a particular value $\alpha_{0,\mbox{\scriptsize init.}}^{\star}$. Using the property~(\ref{F_invariance}) we get
%
%
\begin{multline}
%
\Big\langle
\sum_{\alpha_{0} \neq \alpha_{0,\mbox{\scriptsize init.}}(\sigma)}
\!\!\!\!
\Big\langle F^{\{\alpha\}}_{\alpha_{0}\alpha_{0,\mbox{\scriptsize init.}}(\sigma)} \Big\rangle_{\hat{V}} \times
A_{\vec{n}_{\alpha_{0}^{}}(\sigma)}
\Big\rangle_{\sigma | \alpha_{0,\mbox{\scriptsize init.}}(\sigma) = \alpha_{0,\mbox{\scriptsize init.}}^{\star}}
\!\!\!
=
\\
\sum_{\alpha_{0} \neq \alpha_{0,\mbox{\scriptsize init.}}^{\star}}
\Big\langle F^{\{\alpha\}}_{\alpha_{0}^{}\alpha_{0,\mbox{\scriptsize init.}}^{\star}} \Big\rangle_{\hat{V}}
\times
\label{SUPP_A_relax_4}
\Big\{
\MC{A} - (N-1)^{-1} (\init{A} - \MC{A})
\Big\}
%
\end{multline}
%
%
where
\begin{eqnarray*}
\MC{A} \equiv N^{-1} \sum_{\vec{n}:\,E_{\vec{n}} \in {\cal W}_{\mbox{\scriptsize MC}}(E,\,\Delta E)}^{N} A_{\vec{n}}
\end{eqnarray*}
is the \textit{microcanonical expectation value of the observable}. Notice that $\MC{A}$ depends neither on permutation $\sigma$ nor on perturbation $\hat{V}$. Note that if we were to retain the $\alpha_{0} = \alpha_{0,\mbox{\scriptsize init.}}^{\star}$ term on the left hand side of the equation~(\ref{SUPP_A_relax_4}), the factorization into a product of an intitial-state-dependent and intitial-state-independent factors on the right hand side would not be possible; this, in turn, would make the subsequent simplifications impossible as well. This justifies the singling out the first term in Eqn.~(\ref{SUPP_A_relax_1}), which is the term responsible for the memory of the initial conditions.

The second factor in the r.h.s. of equation~(\ref{SUPP_A_relax_4}) does not depend on the initial state index $\alpha_{0,\mbox{\scriptsize init.}}^{\star}$ at all. Averaging of the first factor over $\alpha_{0,\mbox{\scriptsize init.}}^{\star}$ gives
\begin{eqnarray}
&&
\Big\langle
\sum_{\alpha_{0} \neq \alpha_{0,\mbox{\scriptsize init.}}^{\star}}
\Big\langle F^{\{\alpha\}}_{\alpha_{0}^{}\alpha_{0,\mbox{\scriptsize init.}}^{\star}} \Big\rangle_{\hat{V}}
\Big\rangle_{\alpha_{0,\mbox{\scriptsize init.}}^{\star}}
=
1 - \Big\langle \etaMC(\sigma,\,\hat{V}) \Big\rangle_{\sigma,\,\hat{V}}
\quad.
\label{SUPP_A_relax_5}
\end{eqnarray}

Combining the particular results in equations~(\ref{SUPP_A_relax_3.5},\ref{SUPP_A_relax_4},\ref{SUPP_A_relax_5}), we get
\begin{eqnarray}
&&
\Big\langle \inftave{A} \Big\rangle_{\sigma,\,\hat{V}}
=
\Big\langle \etaMC(\sigma,\,\hat{V}) \Big\rangle_{\sigma,\,\hat{V}} \, \init{A} +
\nonumber
\\
&&
\qquad
\left(1 - \Big\langle \etaMC(\sigma,\,\hat{V}) \Big\rangle_{\sigma,\,\hat{V}}\right) \times
\label{SUPP_almost_the_central_result}
\Big\{
\MC{A} - (N-1)^{-1} (\init{A} - \MC{A})
\Big\}
\\
&&
\hspace*{5em}=
\frac{
(N-\Npca) \init{A}
+
N(\Npca-1) \MC{A}
     }{\Npca(N-1)}
\,\,,
\nonumber
\end{eqnarray}
where the number of the principal components is given by
\begin{eqnarray*}
\Npca \equiv \frac{1}{ \etaMC }
\quad.
\end{eqnarray*}
This concludes the proof for the formula in equation~({\MAINIIalmostIItheIIcentralIIresult}) of the main text.


\begin{thebibliography}{10}
\expandafter\ifx\csname url\endcsname\relax
  \def\url#1{\texttt{#1}}\fi
\expandafter\ifx\csname urlprefix\endcsname\relax\def\urlprefix{URL }\fi
\providecommand{\bibinfo}[2]{#2}
\providecommand{\eprint}[2][]{\url{#2}}

\bibitem{berman2005}
\bibinfo{author}{Berman, G.~P.} \& \bibinfo{author}{Izrailev, F.~M.}
\newblock \bibinfo{title}{The {F}ermi-{P}asta-{U}lam problem: {F}ifty years of
  progress}.
\newblock \emph{\bibinfo{journal}{Chaos}} \textbf{\bibinfo{volume}{15}},
  \bibinfo{pages}{015104} (\bibinfo{year}{2005}).

\bibitem{kinoshita2006}
\bibinfo{author}{Kinoshita, T.}, \bibinfo{author}{Wenger, T.} \&
  \bibinfo{author}{Weiss, D.~S.}
\newblock \bibinfo{title}{A quantum {N}ewton's cradle}.
\newblock \emph{\bibinfo{journal}{Nature}} \textbf{\bibinfo{volume}{440}},
  \bibinfo{pages}{900--903} (\bibinfo{year}{2006}).

\bibitem{hofferberth2007}
\bibinfo{author}{Hofferberth, S.}, \bibinfo{author}{Lesanovsky, I.},
  \bibinfo{author}{Fischer, B.}, \bibinfo{author}{Schumm, T.} \&
  \bibinfo{author}{Schmiedmayer, J.}
\newblock \bibinfo{title}{Non-equilibrium coherence dynamics in one-dimensional
  {B}ose gases}.
\newblock \emph{\bibinfo{journal}{Nature}} \textbf{\bibinfo{volume}{449}},
  \bibinfo{pages}{324--327} (\bibinfo{year}{2007}).

\bibitem{Trotzky2011}
\bibinfo{author}{Trotzky, S.} \emph{et~al.}
\newblock \bibinfo{title}{Probing the relaxation towards equilibrium in an
  isolated strongly correlated 1{D} {B}ose gas}.
\newblock \bibinfo{howpublished}{Preprint at http://arXiv.org/abs/1101.2659
  (2011)}.

\bibitem{rigol2009}
\bibinfo{author}{Rigol, M.}
\newblock \bibinfo{title}{Breakdown of thermalization in finite one-dimensional
  systems}.
\newblock \emph{\bibinfo{journal}{Phys. Rev. Lett.}}
  \textbf{\bibinfo{volume}{103}}, \bibinfo{pages}{100403}
  (\bibinfo{year}{2009}).

\bibitem{reichl1987}
\bibinfo{author}{Reichl, L.~E.} \& \bibinfo{author}{Lin, W.~A.}
\newblock \bibinfo{title}{The search for a quantum {KAM} theorem}.
\newblock \emph{\bibinfo{journal}{Found. Phys.}} \textbf{\bibinfo{volume}{17}},
  \bibinfo{pages}{689--697} (\bibinfo{year}{1987}).

\bibitem{shnirelman1974}
\bibinfo{author}{Shnirelman, A.~I.}
\newblock \bibinfo{title}{Ergodic properties of eigenfunctions}.
\newblock \emph{\bibinfo{journal}{Usp. Mat. Nauk}}
  \textbf{\bibinfo{volume}{29}}, \bibinfo{pages}{181--182}
  (\bibinfo{year}{1974}).

\bibitem{barnett2006}
\bibinfo{author}{Barnett, A.~H.}
\newblock \bibinfo{title}{Asymptotic rate of quantum ergodicity in chaotic
  euclidean billiards}.
\newblock \emph{\bibinfo{journal}{Comm. Pure Appl. Math.}}
  \textbf{\bibinfo{volume}{59}}, \bibinfo{pages}{1457--1488}
  (\bibinfo{year}{2006}).

\bibitem{deutsch1991}
\bibinfo{author}{Deutsch, J.~M.}
\newblock \bibinfo{title}{Quantum statistical mechanics in a closed system}.
\newblock \emph{\bibinfo{journal}{Phys. Rev. A}} \textbf{\bibinfo{volume}{43}},
  \bibinfo{pages}{2046--2049} (\bibinfo{year}{1991}).

\bibitem{srednicki1994}
\bibinfo{author}{Srednicki, M.}
\newblock \bibinfo{title}{Chaos and quantum thermalization}.
\newblock \emph{\bibinfo{journal}{Phys. Rev. E}} \textbf{\bibinfo{volume}{50}},
  \bibinfo{pages}{888--901} (\bibinfo{year}{1994}).

\bibitem{rigol2008}
\bibinfo{author}{Rigol, M.}, \bibinfo{author}{Dunjko, V.} \&
  \bibinfo{author}{Olshanii, M.}
\newblock \bibinfo{title}{Thermalization and its mechanism for generic isolated
  quantum systems}.
\newblock \emph{\bibinfo{journal}{Nature}} \textbf{\bibinfo{volume}{452}},
  \bibinfo{pages}{854--858} (\bibinfo{year}{2008}).

\bibitem{feingold1986}
\bibinfo{author}{Feingold, M.} \& \bibinfo{author}{Peres, A.}
\newblock \bibinfo{title}{Distribution of matrix elements of chaotic systems}.
\newblock \emph{\bibinfo{journal}{Phys. Rev. A}} \textbf{\bibinfo{volume}{34}},
  \bibinfo{pages}{591--595} (\bibinfo{year}{1986}).

\bibitem{horoi1995}
\bibinfo{author}{Horoi, M.}, \bibinfo{author}{Zelevinsky, V.} \&
  \bibinfo{author}{Brown, B.~A.}
\newblock \bibinfo{title}{Chaos vs thermalization in the nuclear shell model}.
\newblock \emph{\bibinfo{journal}{Phys. Rev. Lett.}}
  \textbf{\bibinfo{volume}{74}}, \bibinfo{pages}{5194--5197}
  (\bibinfo{year}{1995}).

\bibitem{flambaum1997}
\bibinfo{author}{Flambaum, V.~V.} \& \bibinfo{author}{Izrailev, F.~M.}
\newblock \bibinfo{title}{Distribution of occupation numbers in finite {F}ermi
  systems and role of interaction in chaos and thermalization}.
\newblock \emph{\bibinfo{journal}{Phys. Rev. E}} \textbf{\bibinfo{volume}{55}},
  \bibinfo{pages}{R13--R16} (\bibinfo{year}{1997}).

\bibitem{georgeot1997}
\bibinfo{author}{Georgeot, B.} \& \bibinfo{author}{Shepelyansky, D.~L.}
\newblock \bibinfo{title}{Breit-{W}igner width and inverse participation ratio
  in finite interacting {F}ermi systems}.
\newblock \emph{\bibinfo{journal}{Phys. Rev. Lett.}}
  \textbf{\bibinfo{volume}{79}}, \bibinfo{pages}{4365--4368}
  (\bibinfo{year}{1997}).

\bibitem{rigol2007}
\bibinfo{author}{Rigol, M.}, \bibinfo{author}{Dunjko, V.},
  \bibinfo{author}{Yurovsky, V.} \& \bibinfo{author}{Olshanii, M.}
\newblock \bibinfo{title}{Relaxation in a completely integrable many-body
  quantum system: An ab initio study of the dynamics of the highly excited
  states of {1D} lattice hard-core bosons}.
\newblock \emph{\bibinfo{journal}{Phys. Rev. Lett.}}
  \textbf{\bibinfo{volume}{98}}, \bibinfo{pages}{050405}
  (\bibinfo{year}{2007}).

\bibitem{calabrese2007}
\bibinfo{author}{Calabrese, P.} \& \bibinfo{author}{Cardy, J.}
\newblock \bibinfo{title}{Quantum quenches in extended systems}.
\newblock \emph{\bibinfo{journal}{J. Stat. Mech.}} \bibinfo{pages}{P06008}
  (\bibinfo{year}{2007}).

\bibitem{cazalilla2006}
\bibinfo{author}{Cazalilla, M.~A.}
\newblock \bibinfo{title}{Effect of suddenly turning on interactions in the
  {L}uttinger model}.
\newblock \emph{\bibinfo{journal}{Phys. Rev. Lett.}}
  \textbf{\bibinfo{volume}{97}}, \bibinfo{pages}{156403}
  (\bibinfo{year}{2006}).

\bibitem{kota2001}
\bibinfo{author}{Kota, V. K.~B.}
\newblock \bibinfo{title}{Embedded random matrix ensembles for complexity and
  chaos in finite interacting particle systems}.
\newblock \emph{\bibinfo{journal}{Phys. Rep.}} \textbf{\bibinfo{volume}{347}},
  \bibinfo{pages}{223--288} (\bibinfo{year}{2001}).

\bibitem{seba1990}
\bibinfo{author}{\v{S}eba, P.}
\newblock \bibinfo{title}{Wave chaos in singular quantum billiard}.
\newblock \emph{\bibinfo{journal}{Phys. Rev. Lett.}}
  \textbf{\bibinfo{volume}{64}}, \bibinfo{pages}{1855--1858}
  (\bibinfo{year}{1990}).

\bibitem{anderson1958}
\bibinfo{author}{Anderson, P.~W.}
\newblock \bibinfo{title}{Absence of diffusion in certain random lattices}.
\newblock \emph{\bibinfo{journal}{Phys. Rev.}} \textbf{\bibinfo{volume}{109}},
  \bibinfo{pages}{1492--1505} (\bibinfo{year}{1958}).

\bibitem{billy2008}
\bibinfo{author}{Billy, J.} \emph{et~al.}
\newblock \bibinfo{title}{Direct observation of {A}nderson localization of
  matter waves in a controlled disorder}.
\newblock \emph{\bibinfo{journal}{Nature}} \textbf{\bibinfo{volume}{453}},
  \bibinfo{pages}{891--894} (\bibinfo{year}{2008}).

\bibitem{roati2008}
\bibinfo{author}{Roati, G.} \emph{et~al.}
\newblock \bibinfo{title}{Anderson localization of a non-interacting
  {B}ose-{E}instein condensate}.
\newblock \emph{\bibinfo{journal}{Nature}} \textbf{\bibinfo{volume}{453}},
  \bibinfo{pages}{895--898} (\bibinfo{year}{2008}).

\bibitem{fishman1982_509}
\bibinfo{author}{Fishman, S.}, \bibinfo{author}{Grempel, D.~R.} \&
  \bibinfo{author}{Prange, R.~E.}
\newblock \bibinfo{title}{Chaos, quantum recurrences, and {A}nderson
  localization}.
\newblock \emph{\bibinfo{journal}{Phys. Rev. Lett.}}
  \textbf{\bibinfo{volume}{49}}, \bibinfo{pages}{509--512}
  (\bibinfo{year}{1982}).

\bibitem{borgonovi1996_4744}
\bibinfo{author}{Borgonovi, F.}, \bibinfo{author}{Casati, G.} \&
  \bibinfo{author}{Li, B.}
\newblock \bibinfo{title}{Diffusion and localization in chaotic billiards}.
\newblock \emph{\bibinfo{journal}{Phys. Rev. Lett.}}
  \textbf{\bibinfo{volume}{77}}, \bibinfo{pages}{4744--4747}
  (\bibinfo{year}{1996}).

\bibitem{altshuler1997_487}
\bibinfo{author}{Altshuler, B.~L.} \& \bibinfo{author}{Levitov, L.~S.}
\newblock \bibinfo{title}{Weak chaos in a quantum {K}epler problem}.
\newblock \emph{\bibinfo{journal}{Phys. Reps.}} \textbf{\bibinfo{volume}{288}},
  \bibinfo{pages}{487 -- 512} (\bibinfo{year}{1997}).

\bibitem{Yurovsky2011}
\bibinfo{author}{Yurovsky, V.~A.} \& \bibinfo{author}{Olshanii, M.}
\newblock \bibinfo{title}{Memory of the initial conditions in an incompletely
  chaotic quantum system: Universal predictions with application to cold
  atoms}.
\newblock \emph{\bibinfo{journal}{Phys. Rev. Lett.}}
  \textbf{\bibinfo{volume}{106}}, \bibinfo{pages}{025303}
  (\bibinfo{year}{2011}).

\bibitem{landau1958}
\bibinfo{author}{Landau, L.~D.} \& \bibinfo{author}{Lifshitz, E.~M.}
\newblock \emph{\bibinfo{title}{Quantum Mechanics: Non-relativistic Theory}}
\newblock  (\bibinfo{publisher}{Butterworth-Heinemann},
  \bibinfo{address}{Oxford, U.K.}, \bibinfo{year}{1991}), pp.
  \bibinfo{pages}{28--29}.

\bibitem{yuzbashyan2002}
\bibinfo{author}{Yuzbashyan, E.~A.}, \bibinfo{author}{Altshuler, B.~L.} \&
  \bibinfo{author}{Shastry, B.~S.}
\newblock \bibinfo{title}{The origin of degeneracies and crossings in the 1d
  {H}ubbard model}.
\newblock \emph{\bibinfo{journal}{J. Phys. A}} \textbf{\bibinfo{volume}{35}},
  \bibinfo{pages}{7525--7547} (\bibinfo{year}{2002}).

\bibitem{berry1977b}
\bibinfo{author}{Berry, M.~V.} \& \bibinfo{author}{Tabor, M.}
\newblock \bibinfo{title}{Level clustering in the regular spectrum}.
\newblock \emph{\bibinfo{journal}{Proc. R. Soc. London, Ser. A}}
  \textbf{\bibinfo{volume}{356}}, \bibinfo{pages}{375--394}
  (\bibinfo{year}{1977}).

\bibitem{cheung1988}
\bibinfo{author}{Cheung, H.-F.}, \bibinfo{author}{Gefen, Y.},
  \bibinfo{author}{Riedel, E.~K.} \& \bibinfo{author}{Shih, W.-H.}
\newblock \bibinfo{title}{Persistent currents in small one-dimensional metal
  rings}.
\newblock \emph{\bibinfo{journal}{Phys. Rev. B}} \textbf{\bibinfo{volume}{37}},
  \bibinfo{pages}{6050--6062} (\bibinfo{year}{1988}).

\bibitem{uski1998}
\bibinfo{author}{Uski, V.}, \bibinfo{author}{Mehlig, B.} \&
  \bibinfo{author}{R{\"o}mer, R.}
\newblock \bibinfo{title}{A numerical study of wave-function and matrix-element
  statistics in the {A}nderson model of localization}.
\newblock \emph{\bibinfo{journal}{Ann. Phys. (Leipzig)}}
  \textbf{\bibinfo{volume}{7}}, \bibinfo{pages}{437--441}
  (\bibinfo{year}{1998}).

\bibitem{heinrichs2009}
\bibinfo{author}{Heinrichs, J.}
\newblock \bibinfo{title}{Absence of localization in a disordered
  one-dimensional ring threaded by an {A}haronov-{B}ohm flux}.
\newblock \emph{\bibinfo{journal}{J. Phys. Cond. Mat.}}
  \textbf{\bibinfo{volume}{21}}, \bibinfo{pages}{295701}
  (\bibinfo{year}{2009}).

\bibitem{seba1991}
\bibinfo{author}{\v{S}eba, P.} \& \bibinfo{author}{\.{Z}yczkowski, K.}
\newblock \bibinfo{title}{Wave chaos in quantized classically nonchaotic
  systems}.
\newblock \emph{\bibinfo{journal}{Phys. Rev. A}} \textbf{\bibinfo{volume}{44}},
  \bibinfo{pages}{3457--3465} (\bibinfo{year}{1991}).

\bibitem{stotland2008}
\bibinfo{author}{Stotland, A.}, \bibinfo{author}{Budoyo, R.},
  \bibinfo{author}{Peer, T.}, \bibinfo{author}{Kottos, T.} \&
  \bibinfo{author}{Cohen, D.}
\newblock \bibinfo{title}{The mesoscopic conductance of disordered rings, its
  random matrix theory and the generalized variable range hopping picture}.
\newblock \emph{\bibinfo{journal}{J. Phys. A}} \textbf{\bibinfo{volume}{41}},
  \bibinfo{pages}{262001} (\bibinfo{year}{2008}).

\bibitem{guhr1998}
\bibinfo{author}{Guhr, T.}, \bibinfo{author}{M{\"u}ller-Groeling, A.} \&
  \bibinfo{author}{Weidenm{\"u}ller, H.~A.}
\newblock \bibinfo{title}{Random-matrix theories in quantum physics: common
  concepts}.
\newblock \emph{\bibinfo{journal}{Phys. Reps.}} \textbf{\bibinfo{volume}{299}},
  \bibinfo{pages}{189--425} (\bibinfo{year}{1998}).

\bibitem{fyodorov1995_R11580}
\bibinfo{author}{Fyodorov, Y.~V.} \& \bibinfo{author}{Mirlin, A.~D.}
\newblock \bibinfo{title}{Statistical properties of random banded matrices with
  strongly fluctuating diagonal elements}.
\newblock \emph{\bibinfo{journal}{Phys. Rev. B}} \textbf{\bibinfo{volume}{52}},
  \bibinfo{pages}{R11580--R11583} (\bibinfo{year}{1995}).

\bibitem{frahm1995_385}
\bibinfo{author}{Frahm, K.} \& \bibinfo{author}{M\"{u}ller-Groeling, A.}
\newblock \bibinfo{title}{Analytical results for random band matrices with
  preferential basis}.
\newblock \emph{\bibinfo{journal}{Europhys. Lett.}}
  \textbf{\bibinfo{volume}{32}}, \bibinfo{pages}{385--390}
  (\bibinfo{year}{1995}).

\bibitem{flambaum1997_5144}
\bibinfo{author}{Flambaum, V.~V.} \& \bibinfo{author}{Izrailev, F.~M.}
\newblock \bibinfo{title}{Statistical theory of finite {F}ermi systems based on
  the structure of chaotic eigenstates}.
\newblock \emph{\bibinfo{journal}{Phys. Rev. E}} \textbf{\bibinfo{volume}{56}},
  \bibinfo{pages}{5144--5159} (\bibinfo{year}{1997}).

\bibitem{wigner1955_548}
\bibinfo{author}{Wigner, E.~P.}
\newblock \bibinfo{title}{Characteristic vectors of bordered matrices with
  infinite dimensions}.
\newblock \emph{\bibinfo{journal}{Ann. Math}} \textbf{\bibinfo{volume}{62}},
  \bibinfo{pages}{548--564} (\bibinfo{year}{1955}).

\bibitem{jacquod1995_3501}
\bibinfo{author}{Jacquod, P.} \& \bibinfo{author}{Shepelyansky, D.~L.}
\newblock \bibinfo{title}{Hidden {B}reit-{W}igner distribution and other
  properties of random matrices with preferential basis}.
\newblock \emph{\bibinfo{journal}{Phys. Rev. Lett.}}
  \textbf{\bibinfo{volume}{75}}, \bibinfo{pages}{3501--3504}
  (\bibinfo{year}{1995}).

\bibitem{manoharan2000}
\bibinfo{author}{Manoharan, H.~C.}, \bibinfo{author}{Lutz, C.~P.} \&
  \bibinfo{author}{Eigler, D.~M.}
\newblock \bibinfo{title}{Quantum mirages: The coherent projection of
  electronic structure}.
\newblock \emph{\bibinfo{journal}{Nature}} \textbf{\bibinfo{volume}{403}},
  \bibinfo{pages}{512--515} (\bibinfo{year}{2000}).

\bibitem{brown2010}
\bibinfo{author}{Brown, W.}
\newblock \emph{\bibinfo{title}{Random Quantum Dynamics: From Random Quantum
  Circuits to Quantum Chaos}}.
\newblock Ph.D. thesis, \bibinfo{school}{Dartmouth College, NH, USA}
  (\bibinfo{year}{2010}).

\bibitem{flambaum1996_5729}
\bibinfo{author}{Flambaum, V.~V.}, \bibinfo{author}{Gribakin, G.~F.} \&
  \bibinfo{author}{Izrailev, F.~M.}
\newblock \bibinfo{title}{Correlations within eigenvectors and transition
  amplitudes in the two-body random interaction model}.
\newblock \emph{\bibinfo{journal}{Phys. Rev. E}} \textbf{\bibinfo{volume}{53}},
  \bibinfo{pages}{5729--5741} (\bibinfo{year}{1996}).

\bibitem{valiente2011}
\bibinfo{author}{Valiente, M.} \& \bibinfo{author}{M{\o}lmer, K.}
\newblock \bibinfo{title}{Quasi-one-dimensional scattering in a discrete
  model}.
\newblock \bibinfo{howpublished}{Preprint at http://arXiv.org/abs/1107.5459
  (2011)}.

\bibitem{Sayed2008}
\bibinfo{author}{Sayed, A.~H.}
\newblock \emph{\bibinfo{title}{Adaptive Filters}}
\newblock  (\bibinfo{publisher}{Wiley}, \bibinfo{address}{Hoboken, N.J.},
  \bibinfo{year}{2008}), pp. \bibinfo{pages}{10--11}.

\end{thebibliography}
\end{document}